\documentclass[pra,twocolumn,preprintnumbers,floatfix]{revtex4}
\usepackage{float}
\usepackage{graphicx}
\usepackage{dcolumn}
\usepackage{bm}
\usepackage{latexsym,epsfig}
\usepackage{graphicx}
\usepackage{verbatim}
\usepackage{comment}
\usepackage{amsmath}
\usepackage{amssymb}
\usepackage{stmaryrd}
\usepackage{color}
\usepackage{epstopdf}
\usepackage{grffile}
\usepackage{hyperref}
\hypersetup{
    colorlinks=true,      
    urlcolor=blue,
    citecolor=blue,
    linkcolor=blue
}
\DeclareGraphicsExtensions{.eps}
\newcommand{\beq}{\begin{equation}}
\newcommand{\eeq}{\end{equation}}
\newcommand{\bea}{\begin{eqnarray}}
\newcommand{\eea}{\end{eqnarray}}
\newcommand{\ben}{\begin{eqnarray*}}
\newcommand{\een}{\end{eqnarray*}}
\newcommand{\bfig}{\begin{figure}}
\newcommand{\efig}{\end{figure}}

\begin{document}
\title{Quantum phases of constrained bosons on a two-leg Bose-Hubbard ladder}
\author{Ashirbad Padhan$^1$, Rajashri Parida$^{2,3}$, Sayan Lahiri$^1$, Mrinal Kanti Giri$^{4}$ and Tapan Mishra$^{2,3}$}
\email{mishratapan@gmail.com}
\affiliation{$^1$Department of Physics, Indian Institute of Technology, Guwahati-781039, India\\
$^2$School of Physical Sciences, National Institute of Science Education and Research, Jatni 752050, India\\
$^3$Homi Bhabha National Institute, Training School Complex, Anushaktinagar, Mumbai 400094, India\\
$^4$Centre for Quantum Engineering, Research and Education, TCG CREST, Salt Lake, Kolkata 700091, India}
\date{\today}

\begin{abstract}
Bosons in periodic potentials with very strong local interactions, known as the constrained bosons often exhibit interesting physical behavior. We investigate the ground state properties of a two-leg Bose-Hubbard ladder by imposing three-body constraint in one leg and hardcore constraint in the other. By using the cluster-mean-field theory approximation and the density matrix renormalization group method, we show that at unit filling, for strong two-body attraction among the three-body constrained bosons, the system  becomes a gapped pair-Mott insulator where all the bosons form strong bound pairs and occupy the leg with three-body constraint. With increase in hopping strength this pair-Mott insulator phase undergoes a phase transition to the gapless superfluid phase for equal leg and rung hopping strengths. However, when the rung hopping is stronger compared to the leg hopping, we obtain a crossover to another gapped phase which is called the rung-Mott insulator phase where the bosons prefer to delocalize on the rungs than the legs. By moving away from unit filling, the system remains in the superfluid phase except for a small region below the gapped phase where a pair superfluid phase is stabilized in the regime of strong attractive interaction. We further extend our studies by considering three-body constraint on both the legs and find that the crossover from the gapped to gapped phase does not occur rather the system undergoes a transition from a pair-rung-Mott insulator phase to the superfluid phase at unit filling. Moreover, in this case we find the signature of the pair superfluid phase on either sides of this gapped phase.
\end{abstract}

\maketitle

\section{Introduction}\label{introduction}

Strongly correlated bosonic systems serve as a very active playground for the realization of novel and exotic quantum phases of matter. These systems have gained considerable interest in the last few decades, mainly motivated by the spectacular manipulation of ultracold atoms in optical lattices~\cite{lewenstein2007}. Starting from the path breaking observation of the superfluid (SF) to Mott insulator (MI) transition in such systems~\cite{greiner2002}, a plethora of interesting studies have been performed in recent years leading to various new directions in condensed matter physics, AMO physics and quantum technologies~\cite{esslinger2010fermi,tarruell2018quantum,dutta2015non, kennett2013out}. Due to the versatility in terms of flexibility in tuning system parameters and geometry, the systems of ultracold atoms in optical lattices have been widely used for quantum simulations of strongly correlated systems~\cite{bloch2005exploring, bloch2012ultracold, bloch2008many,gross2017quantum,schafer2020tools,jin2022manipulation,lewenstein2007}. On the other hand, several other quantum simulators based on arrays of trapped ions, superconducting circuits, optical cavities and photonic lattices have been built in recent years to address many-body bosonic systems~\cite{hartmann2016quantum,luo2015quantum,blatt2012quantum,liu2014optical,monroe2021programmable,bollinger2013simulating,aspuru2012photonic}. 

In this context, the low-dimensional bosonic systems, in particular one-dimensional (1D) and quasi-1D systems are a topic of paramount interest of research owing to the dominant role played by the interactions and strong correlation~\cite{cazalilla2011one}. Among various low dimensional lattice models, the two-leg ladder systems are  particularly important because they are intermediate to the one and two dimensional lattice system and can provide insights about the physics while shifting from one to two dimensional lattices~\cite{dagotto1996surprises}. 

One of the simplest such models is the two-leg Bose-Hubbard (BH) ladder which has been widely studied theoretically and has been experimentally simulated in various artificial systems~\cite{atala2014observation,greschner2015spontaneous,giamarchi2016current,tokuno2014ground,halati2017cavity,uchino2015population,kelecs2015mott,tschischik2012nonequilibrium,dhar2012bose,takayoshi2013phase,ahlbrecht2012,zhu2022observation,ye2019propagation}. Starting from the study of combined effect of onsite interactions and rung-hopping on the SF-MI transition~\cite{donohue2001, luthra2008} a wealth of novel physics has been discovered in presence of superlattice potential~\cite{danshita2007quantum,danshita2008reentrant,lahiri2020mott}, longer range interaction~\cite{berg2008rise,sachdeva2017extended},  geometric frustration ~\cite{dhar2013chiral,orignac2001meissner,halati2023bose,barbiero2019coupling,an2017direct,sachdeva2018two,piraud2015vortex,greschner2016symmetry,greschner2017vortex,mishra2015polar,greschner2018quantum,PhysRevB.87.174504,dhar2011,dhar2012bose,qiao2021quantum,haller2020exploring,cceven2022neural,PhysRevA.91.013629,aidelsburger2015artificial}, lattice topology~\cite{PhysRevA.89.023619,mondal2021topological,barbiero2018quenched,berg2008rise} and disorder~\cite{orignac1998vortices,PhysRevB.101.134203,carrasquilla2011} in systems of BH ladder. 

Bosons on a two-leg ladder subjected to various constraints such as the hardcore (maximum occupation of one boson per site) or three-body constraint (maximum occupation of two bosons per site) with only local and/or short range interactions are known to exhibit interesting physics~\cite{cazalilla2011}. These systems carry enough significance as they can be appropriately mapped to various spin models and variants of the Hubbard model to mimic their physics in a bosonic platforms and hence paving path for the experimental realization. One of the interesting manifestations of the hardcore bosons (HCBs) on a two-leg ladder is the appearance of the gapped rung-Mott insulator (RMI) phase for any finite rung-hopping at half filling where in each rung a hardcore boson delocalizes to form a rung-singlet~\cite{crepin2011}. Addition of nearest neighbor interactions along the legs and rungs of the ladder are known to stabilizes the charge density wave (CDW) and the supersolid (SS) phases~\cite{pandey2017triplet}. 
A study by some of us has also uncovered the existence of a dimer rung-insulator (DRI) phase at unit filling in a system of three-body constrained bosons on a two-leg ladder in presence of attractive onsite interaction and repulsive nearest-neighbor interaction along the rungs of the ladder~\cite{singh2014}. In the DRI phase, strongly bound pairs of bosons delocalize on the rungs of the ladder creating a situation similar to the hardcore bosonic ladder at half filling. A system of two- and three-body constrained bosons in a pair of one-dimensional lattices coupled to each other by non-local attractive interactions exhibits the trimer and dimer superfluid phases with the bosons possessing repulsive onsite interactions~\cite{singh2017}.

\begin{figure}[t]
\centering
\includegraphics[width=0.70\linewidth]{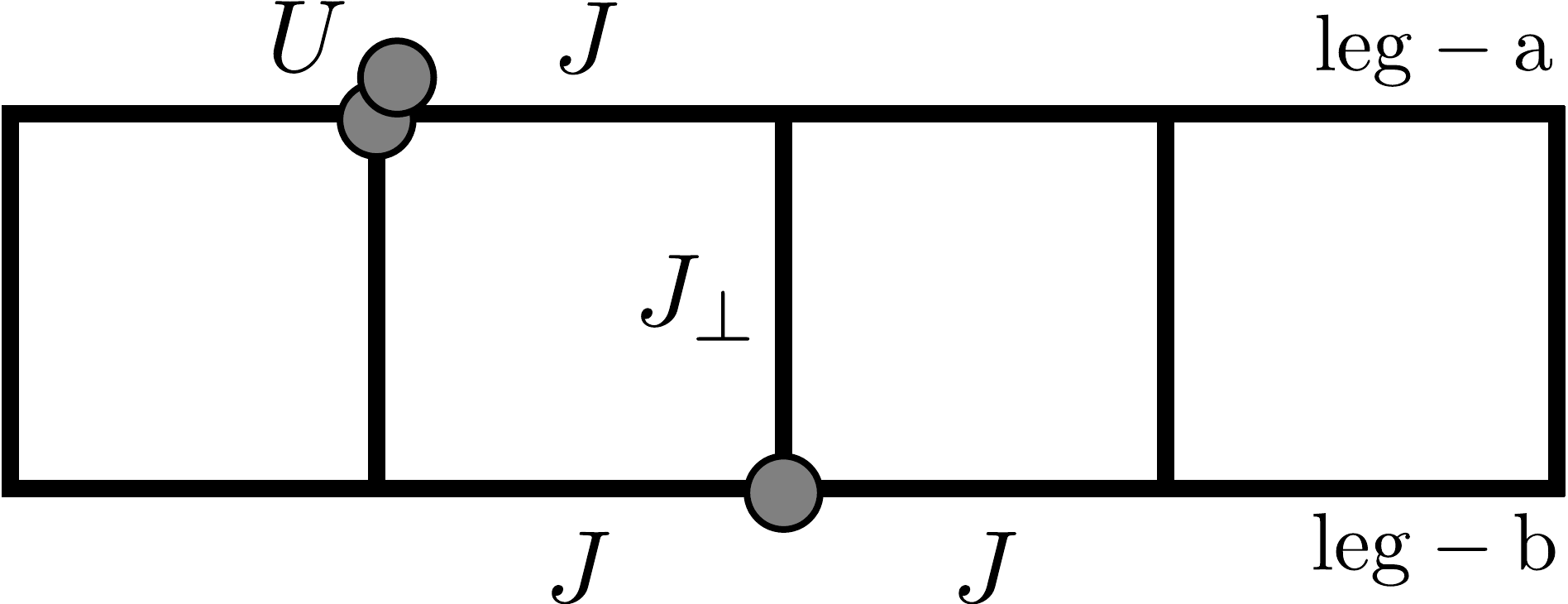}
\caption{ Schematic diagram of a two-leg Bose-Hubbard ladder with leg-hopping $J$, rung-hopping $J_{\perp}$ and onsite interaction $U$.}
\label{fig:bh-ladder}
\end{figure}

However, the ground state properties of a two-leg ladder with two different types of constraints imposed on the legs is not well explored. 
In this paper we consider a two-leg BH ladder as depicted in Fig.~\ref{fig:bh-ladder} and impose  three-body constraint (TBC) in one leg and hardcore constraint (HC) in the other leg. By allowing attractive onsite  interactions for the bosons residing in the leg having the TBC, we study the combined role of onsite interaction and rung-to-leg hopping ratio on the ground state properties of the system. Using an approximation method and a sophisticated numerical method we reveal that when the rung-to-leg ratio is small, an MI phase of bosonic pairs or the pair-MI (PMI) phase is formed on the leg having the TBC imposed which undergoes a transition to the SF phase as a function of interaction. On the other hand, in the limit of large rung-to-leg hopping ratio, the ground state phase diagram exhibits a crossover from the PMI phase to rung-MI (RMI) phase at unit filling as a function of the attractive interaction.  Moreover, we find the signatures of the pair superfluid (PSF) phase by moving away from the commensurability in the limit of strong interactions. In the end we compare our results by considering TBC on both the legs where the RMI phase does not appear at unit filling. Rather, we obtain a phase transition from a pair rung-Mott insulator (PRMI) phase to the SF phase at unit filling. In the following we discuss our results in details. 

The structure of the paper is as follows. In Sec.~\ref{modelmethod} we give the details of the system under consideration, the model Hamiltonian and the numerical methods used. Sec.~\ref{results} is devoted to the results and discussions where we first discuss the TBC-HC system and then the TBC-TBC system. Finally, we summarize our results in Sec.~\ref{conc}.

\section{Model and method}\label{modelmethod}

The two-leg BH ladder can be described by the Hamiltonian
\begin{eqnarray}
 H_{BH}&=&-J\sum\limits_{\langle i,j\rangle,\alpha}({a}^\dagger_{i\alpha}{a}_{j\alpha} + H.c.)-J_{\perp}\sum\limits_{i}({a}^\dagger_{ia}{a}_{ib} + H.c.)\nonumber\\
 &+&\sum\limits_{i,\alpha}\frac{U_{\alpha}}{2}{n}_{i\alpha}({n}_{i\alpha} -1)- \sum\limits_{i,\alpha}\mu_{\alpha}{n}_{i\alpha}.
 \label{eq:ham5}
\end{eqnarray}
Here ${a}^\dagger_{i\alpha}({a}_{i\alpha})$ is the bosonic creation (annihilation) operator at the $i^{th}$ site of the $\alpha^{th}$ leg, where $\alpha$ $(=a, b)$ represents the leg index of the ladder. $U_{\alpha}$ represents the two-body onsite interactions and $\mu_{\alpha}$ is the chemical potentials on the legs. $J$ and $J_{\perp}$ are the hopping strengths of bosons along the legs and rungs of the ladder respectively. ${n}_{i\alpha}$ is the number operator corresponding to the $i^{th}$ site in the $\alpha^{th}$ leg. First we impose three-body constraint  in leg-a and hardcore constraint in leg-b which are achieved by considering  $(a^{\dagger}_a)^3=0$ and $(a^{\dagger}_b)^2=0$, respectively. It is to be noted that for the HC in leg-b i.e. $U_{\alpha=b}\rightarrow \infty$ the terms associated to $U_{\alpha=b}$ vanish in Eq.~(\ref{eq:ham5}). However, due to the TBC in leg-a, $U_{\alpha=a}$ remains finite. Hereafter, we denote $U_{\alpha=a}$ as $U$ without loss of generality. 

For the exploration of the ground state properties of the model shown in Eq.~\ref{eq:ham5}, we first employ the cluster-mean-field theory (CMFT) approach~\cite{mcintosh2012multisite,yamamoto2012quantum,huerga2013composite,hassan2010slave,singh2017quantum,singh2014quantum} and then the density matrix renormalization group (DMRG) method based on the matrix product state (MPS) approach~\cite{white1992,schollwock2005density, schollwock2011density,cirac2021matrix}. Note that the CMFT method is based on the mean-field approximation and owing to its construction, this method provides a qualitative understanding of the underlying physics with less computing effort. However, the DMRG method is well suited to accurately obtain the physics of the low-dimensional systems such as the one considered here.  

Using the CMFT approach we can write 
\begin{equation}
 {H}_{BH}={H}_{C}+{H}_{MF},
 \label{eqn:eq2}
\end{equation}
where, $H_C~(H_{MF})$ is the cluster~(mean-field) part of the Hamiltonian. $H_C$ is same as Eq.~\ref{eq:ham5} limited to the cluster size and is treated exactly. 
Introducing the leg-dependent SF order parameter and the superfluid density given by 
\begin{equation}
\psi_{i\alpha}={\langle{a}^\dagger_{i\alpha}\rangle}={\langle{a}_{i\alpha}\rangle}
\end{equation}
and 
\begin{equation}
\rho_s=\frac{1}{4}\sum\limits_{i=1}^{2}\sum\limits_{\alpha \in[a,b]}|\psi_{i\alpha}|^2
\end{equation}
respectively, we write ${H}_{MF}$ as 
\begin{eqnarray}
 {H}_{MF} &=& -J\sum\limits_{\langle i,j \rangle,\alpha}[(a^\dagger_{i\alpha}+a_{i\alpha})\psi_{j\alpha}-\psi_{i\alpha}^* \psi_{j\alpha}].
\end{eqnarray}
We also assume equal chemical potentials for bosons in both the legs by making $\mu_a=\mu_b=\mu$ and use a $4-$ sites cluster for the CMFT calculations. 

The DMRG simulations are performed in canonical ensemble with a fixed boson number and hence the Hamiltonian in Eq.~\ref{eq:ham5} is explicitly independent of $\mu$. We apply the MPS based DMRG algorithm using an open boundary condition on a system of $L$ sites which is equivalent to $L/2$ rungs. In our simulations, we have considered up to $L=200$ and bond dimensions up to $400$. Our studies are focused on attractive onsite interaction for the bosons in the leg-a. To ensure an attractive onsite interaction between the particles the results are obtained by considering $U=-1$ which also sets the energy scale that makes all the physical parameters dimensionless.

\section{Results and Discussion}\label{results}
In this section we discuss the results in detail for the TBC-HC system, first from the CMFT analysis and then from the DMRG analysis. Then we briefly discuss the TBC-TBC phase diagrams for comparison. 

\subsection{The TBC-HC system}\label{hcbtbc}
\subsubsection{The CMFT phase diagram}
\begin{figure}[!t]
\centering
\includegraphics[width=1\linewidth]{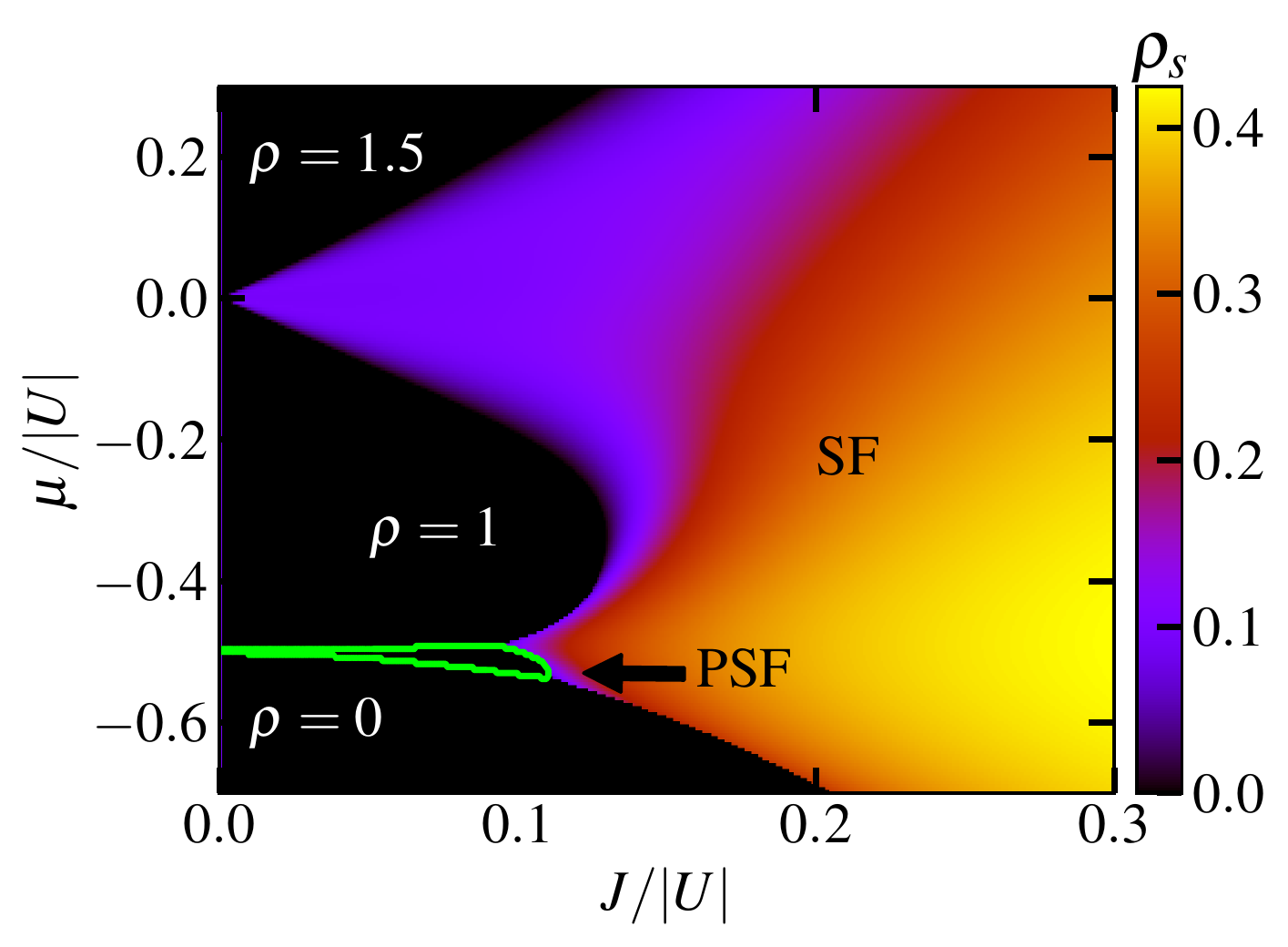}
\caption{Phase diagram of the TBC-HC system in the $J/|U|-\mu/|U|$ plane using the CMFT approach for $J_{\perp}/J=1$ on a $4$-site cluster. The black (colored) region is the gapped (gapless) region and the solid green line denotes the boundary of the PSF phase. The color bar represents the values of the superfluid density $\rho_s$.}
\label{fig:hcbtbc_bhmodel1}
\end{figure}

\begin{figure}[t]
\centering
\includegraphics[width=1\linewidth]{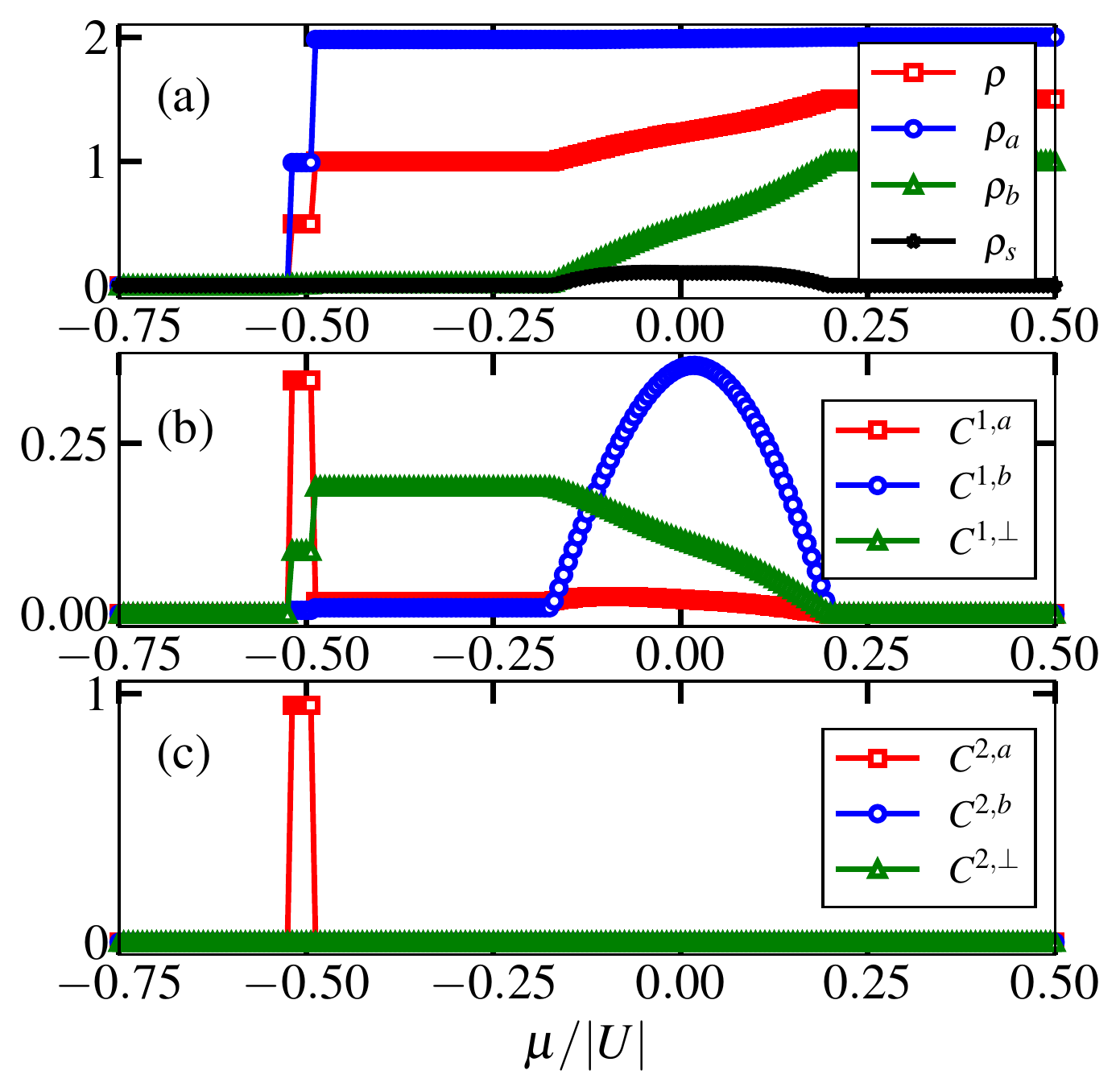}
\caption{(a) $\rho$, $\rho_a$, $\rho_b$ and $\rho_s$ are plotted as a function of $\mu/|U|$. (b) and (c) depict the CMFT data for the nearest-neighbor single- and pair-correlations (as defined in Eq.~\ref{eq:corr5}), respectively as a function of $\mu/|U|$ for $J/|U|=0.09$ and $J_{\perp}/J=1$.}
\label{fig:hcbtbc-rhomu}
\end{figure}

We begin by depicting the phase diagram in Fig.~\ref{fig:hcbtbc_bhmodel1} which is obtained by using the CMFT approach for $J_\perp/J=1$ for the TBC-HC case where we plot the superfluid density $\rho_s$ in the $J/|U|-\mu/|U|$ plane. This  exhibits an insulating lobe  at $\rho=1$ (black region where $\rho_s=0$) surrounded by the gapless  regions (light colored regions where $\rho_s\neq 0$). Now looking at the behavior of different physical quantities within the reach of the CMFT approach we identify different quantum phases which are depicted in the phase diagram of Fig.~\ref{fig:hcbtbc_bhmodel1}. The gapped and gapless phases are identified from the behavior of $\rho_s$ (black curve) and the total density $\rho=\rho_a+\rho_b$ (red curve) as a function of $\mu/|U|$ as shown in Fig.~\ref{fig:hcbtbc-rhomu}(a) for an exemplary cut through the phase diagram at $J/|U|=0.09$. The finite plateau and vanishing of $\rho_s$ at $\rho=1$ for a range of values of $\mu/|U|$ between $-0.4875$ and $-0.175$ indicate the gapped region in the phase diagram. Moving away from $\rho=1$, we find that while $\rho$ continuously increases as a function of $\mu$ in the region above the plateau, the change in $\rho$ occurs in steps that corresponds to a change of two particles at a time in the region below the plateau. Within our CMFT approach, these discrete jumps along with vanishing $\rho_s$ in the latter case are the characteristics of a PSF phase and the former is the indication of the SF phase~\cite{singh2014,singh2016,singh2017}. The PSF region appears in the regime of small $J/|U|$ values which is demarcated by the green boundary and is sandwiched between the $\rho=1$ and $\rho=0$ regions in Fig.~\ref{fig:hcbtbc_bhmodel1}. The gapless SF (PSF) phase extends up to $\rho=1.5$ ($\rho=0$) limits starting from the $\rho=1$ lobe as can be seen from Fig.~\ref{fig:hcbtbc-rhomu}(a). Note that the black regions above and below the $\rho=1$ lobe are the full ($\rho=1.5$) and empty ($\rho=0$) states, respectively.
In the following we characterize these phases in detail.

We first examine the behavior of the individual leg densities i.e. $\rho_a$ and0.18 $\rho_b$ in Fig.~\ref{fig:hcbtbc-rhomu}(a). It can be seen that, as $\mu$ increases and the plateau region at $\rho=1$ is reached, the leg-a gets populated first and the density of the leg-a reaches a value $\rho_a\sim 2$ (blue circles). However, the density of leg-b remains close to zero i.e. $\rho_b\sim 0$ (green triangles). In this case the occupation of leg-a is favored because of the three-body constraint and attractive nature of the interaction as compared to the hardcore constraint in leg-b. With further increase in $\mu$, the density of the system $\rho$ increases and becomes incommensurate. In this regime, $\rho_b$ starts to increase and eventually reaches its maximum value at $\rho_b=1$ and at the same time $\rho_a$ saturates to  the value $\rho_a=2$. At this saturation point the total density of the system, $\rho=1.5$ which is the full state. To further understand these phases, we compute various correlation functions within the cluster such as 
\begin{eqnarray}
C^{n,a}=\langle (a_{ia}^{\dagger})^n (a_{ja})^n\rangle\nonumber\\
C^{n,b}=\langle (a_{ib}^{\dagger})^n (a_{jb})^n\rangle\nonumber\\
C^{n,\perp}=\langle (a_{ia}^{\dagger})^n (a_{ib})^n\rangle.
\label{eq:corr5}
\end{eqnarray}
Here $n=1$ and $n=2$ represent the single and pair correlation functions, respectively. The superscripts $a,~b$ and $\perp$ represent the correlation functions along leg-a, leg-b and along the rung of the ladder, respectively. Note that due to finite size of the cluster considered here (i.e. four sites), only the nearest-neighbor (NN) correlation functions are computed.

In Fig.~\ref{fig:hcbtbc-rhomu}(b) and Fig.~\ref{fig:hcbtbc-rhomu}(c), we plot the single and pair correlation functions as a function of $\mu/|U|$ for $J/|U|=0.09$. It can be seen from Fig.~\ref{fig:hcbtbc-rhomu}(b) that when the system is gapped (plateau region at $\rho=1$ in Fig.~\ref{fig:hcbtbc-rhomu}(a)), $C^{1,a}$ and $C^{1,b}$ are vanishingly small, whereas $C^{1,\perp}$ remains finite. These features indicate that the motion of the particles along the legs are seized due to almost full and almost empty states of leg-a and leg-b respectively at very small value of $J/|U|$. On the other hand, all the single particle correlations remain finite in the SF phase although  $C^{1,a}$ is comparatively smaller than the other two (Fig.~\ref{fig:hcbtbc-rhomu}(b)). The reason behind this is the following. When in the SF phase, due to the TBC imposed on the particles, the leg-a gets populated first leading to a larger value of $\rho_a$ as compared to $\rho_b$, as can be seen from Fig.~\ref{fig:hcbtbc-rhomu}(a). Hence, the particles' motion in leg-a is restricted leading to small values of $C^{1,a}$. However, when the system is in the PSF phase, the $C^{1,b}$ vanishes whereas the other two correlations remain finite. This is because of the TBC, all the particles prefer to populate leg-a in the regime of small $\mu/|U|$ as compared to leg-b where HC is imposed. Interestingly, all the pair-correlations such as $C^{2,a}$, $C^{2,b}$ and $C^{2,\perp}$ vanish in the gapped and gapless phases except in the PSF phase where the $C^{2,a}$ remains finite  in the regime $-0.525<\mu/|U|<-0.4875$. The formation of the PSF phase in this regime is due to the attractive nature of $U$ for which  the particles in leg-a form bound pairs. 
\begin{figure}[!t]
\centering
\includegraphics[width=1\linewidth]{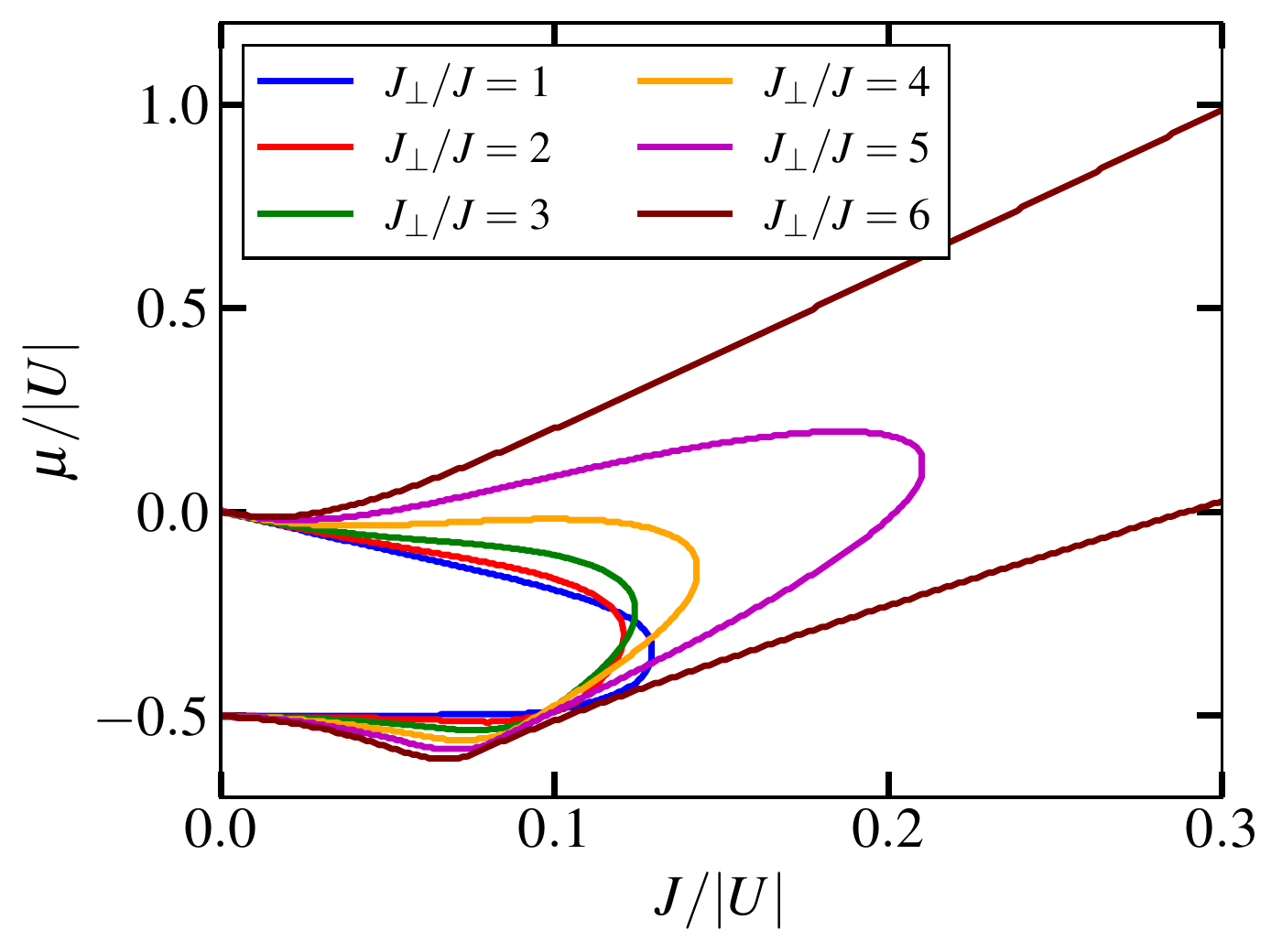}
\caption{CMFT phase diagram of the TBC-HC ladder in $J/|U|-\mu/|U|$ plane for different $J_{\perp}/J$ ratios. Here the curves represent the boundary of the gapped phase(s) at $\rho=1$.}
\label{fig:hcbtbc_bhmodel3}
\end{figure}

\begin{figure}[!b]
\centering
\includegraphics[width=1\linewidth]{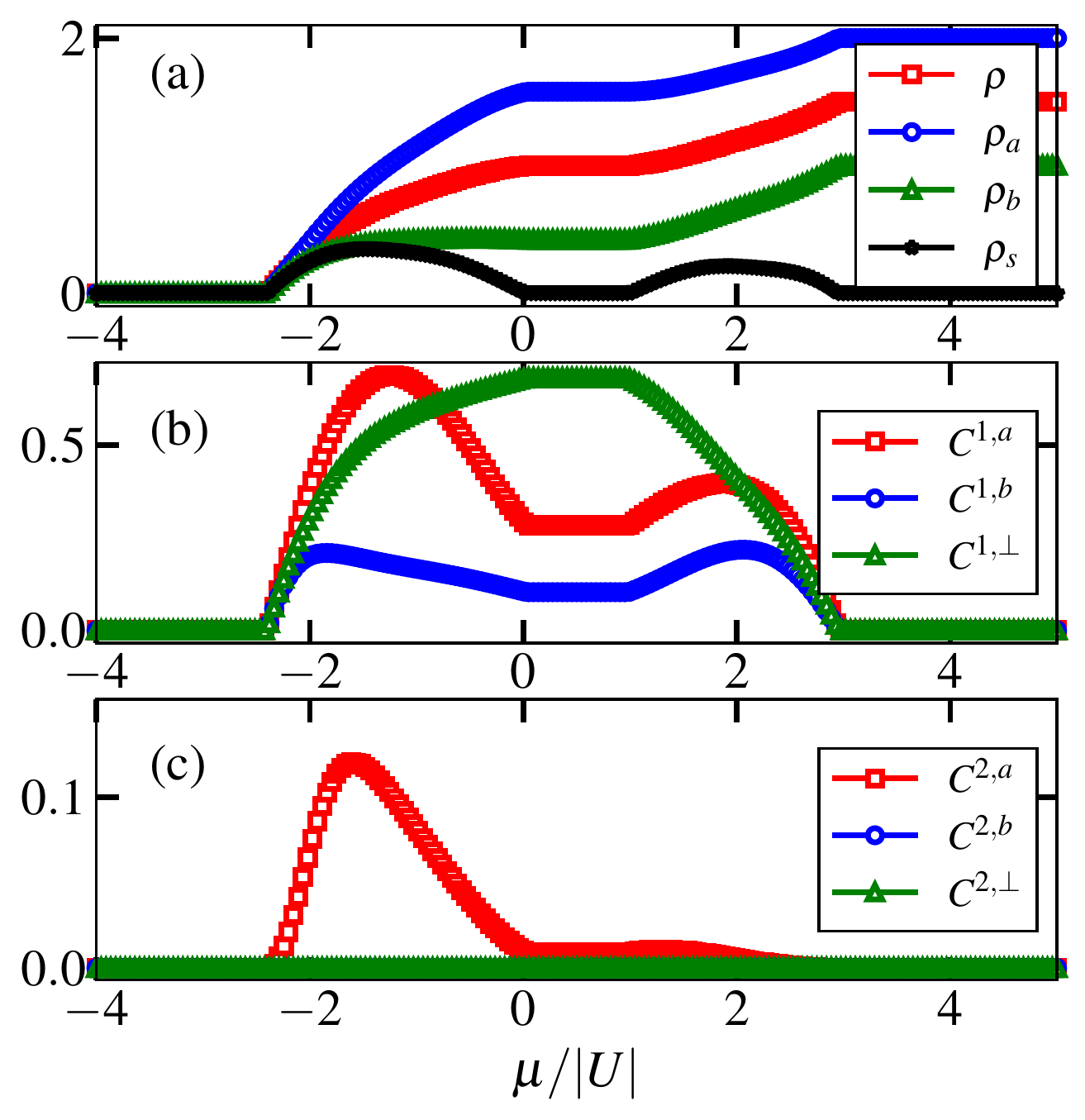}
\caption{(a) $\rho$, $\rho_a$, $\rho_b$ and $\rho_s$ are plotted as a function of $\mu/|U|$. (b) and (c) depict the CMFT data for the nearest-neighbor single- and pair-correlations (as defined in Eq.~\ref{eq:corr5}), respectively as a function of $\mu/|U|$ for $J/|U|=0.3$ and $J_{\perp}/J=3$.}

\label{fig:hcbtbc-rhomu2}
\end{figure}

We now examine the effect of $J_{\perp}/J$ on the gapped phase at $\rho=1$. In Fig.~\ref{fig:hcbtbc_bhmodel3} we plot only the gapped lobes in the $J/|U|-\mu/|U|$ plane at $\rho=1$ for different values of $J_{\perp}/J$ ($=1,~2,~3,~4,~5$ and $6$). With increase in $J_{\perp}/J$,  we find that the gapped lobes gradually expand and eventually for $J_{\perp}/J=6$, the gap does not close at all. In this limit the system remains gapped as a function of $J/|U|$ which indicates a ceasing of the gapped-gapless transition after a certain $J_{\perp}/J$ ratio. Now in order to understand the emergence of gap in the limit of lare $J/|U|$ for $J_{\perp}/J=6$, we look at the behavior of $\rho$ (red square), $\rho_s$ (black star), $\rho_a$ (blue circle), $\rho_b$ (green triangle)  as a function of $\mu$ along the cut at $J/|U|=0.3$. From Fig.~\ref{fig:hcbtbc-rhomu2}, we identify the plateau at $\rho=1$ ($\rho_s=0$) as the gapped phase. Compared to the small $J/|U|$ (or strong $|U|$) limit, here we see a completely different feature. In this case, both the legs start getting populated simultaneously after $\mu/|U|\sim-2.5$ due to the dominant hopping strength (or weak onsite interaction) and also $\rho_a > \rho_b$ for the entire range of $\mu$ due to the constraints imposed. In the plateau region,  the individual leg densities $\rho_a\sim 1.5$ and $\rho_b \sim 0.5$ (see Fig.~\ref{fig:hcbtbc-rhomu2}(a)). We also see a continuous change in $\rho_a$ in the shoulder regions which rules out the presence of any PSF phase. Further, we plot the correlation function in Fig.~\ref{fig:hcbtbc-rhomu2} which shows that $C^{1,\perp}$  dominates over both $C^{1,a}$ and $C^{1,b}$ in the plateau region. The strong $C^{1,\perp}$ in the plateau region (i.e. the gapped region) is the indication of an RMI phase of hardcore bosons. However, in the gapped region, we see a finite single particle correlation in both the legs which is not expected when the system is gapped. On the other hand, the pair-correlations $C^{2,b}$ and $C^{2,\perp}$ vanish throughout and $C^{2,a}$ remains finite though small due to the second order hopping processes in the gapless region. (see Fig.~\ref{fig:hcbtbc-rhomu2} (c)). 

From the above CMFT analysis it is inferred that in the limit of small $J_{\perp}/J$, a phase transition from a gapped to gapless phase occurs when $\rho=1$ as a function of $J/|U|$. However, for large $J_{\perp}/J$, the gap remains finite throughout as a function of $J/|U|$ . Although the gapped phases at small and large $J/|U|$ values exhibit finite rung correlations, for the latter case the rung correlation is strong and the leg correlations remain finite. Moreover, we find the signatures of a PSF phase below the gapped lobe when $J/|U|$ is small. 

It is to be noted that within the CMFT approach it is difficult to clearly identify the nature of these phases and therefore, in the following we employ the DMRG method to concretely establish these quantum phases.

\subsubsection{The DMRG phase diagram}
\begin{figure}[!t]
\centering
\includegraphics[width=1\linewidth]{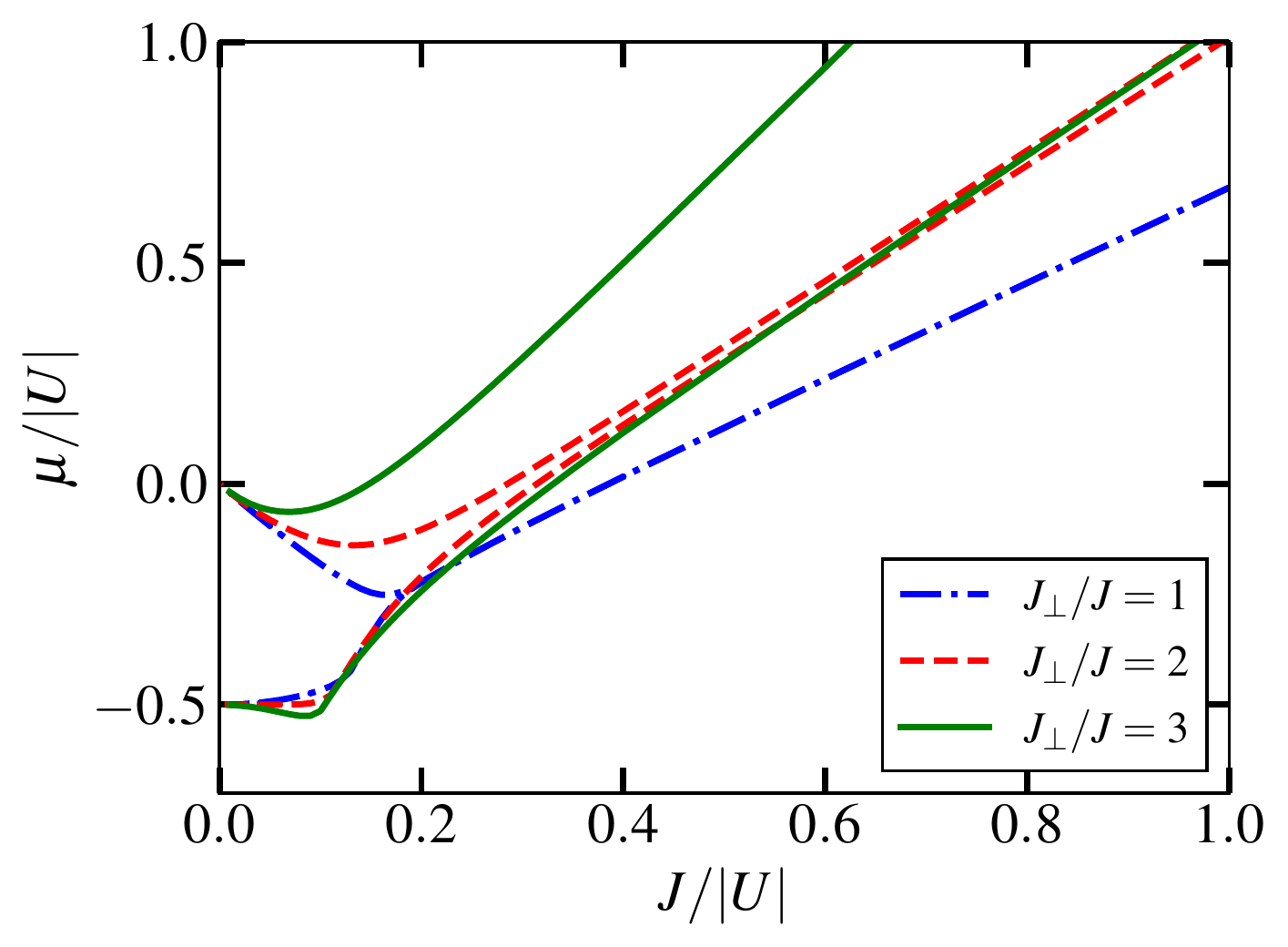}
\caption{DMRG data showing the gapped regions for the TBC-HC ladder at $\rho=1$ for $J_{\perp}/J= 1$ (blue dot-dashed lines), $2$ (red dashed lines) and $3$ (green solid lines). The boundaries represent the extrapolated values of $\mu^+$ and $\mu^-$ with $L=80, 120, 160$ and $200$.}
\label{fig:hcbtbc_phases_varying}
\end{figure}

In this part of the paper we discuss the analysis based on the DMRG simulations of the model shown in Eq.~\ref{eq:ham5} for TBC-HC system. The primary focus of our DMRG simulation is to examine the nature of these gapped phases that arise at $\rho=1$ from the CMFT analysis. To this end we first obtain the gapped regions in the $J/|U| -~ \mu/|U|$ plane for different values of $J_\perp/J$ such as $J_\perp/J=1$ (blue dot-dashed lines), $J_\perp/J=2$ (red dashed lines) and $J_\perp/J=3$ (green solid lines) in Fig.~\ref{fig:hcbtbc_phases_varying}. 
The boundaries of these regions are obtained by computing the chemical potentials as 
\begin{equation}
 \mu^+=(E_{N+2}-E_N)/2 ~\rm{and} ~\mu^-=(E_N-E_{N-2})/2
\end{equation}
where $E_N$ is the ground state energy of the system with $N$ bosons and the gap is defined as $G=\mu^+-\mu^-$. Here we consider two-particle excitation energies in the calculation because of the possibility of boson pair excitation around the gapped region due to the attractive nature of the interaction $U$.
Similar to the CMFT results, we obtain that when $J_{\perp}/J=1$ (blue dot-dashed lines), a transition from a gapped to gapless phase occurs after a  critical $J/|U|\sim 0.19$. However, the situation becomes interesting when $J_{\perp}/J=2$ where the gap decreases, attains its minimum value  near $J/|U|\sim 0.6$ and increases again for higher $J/|U|$ values. This behavior is more prominent for the case of $J_\perp/J=3$ (green solid curve). 
\begin{figure}[!t]
\centering
\includegraphics[width=1\linewidth]{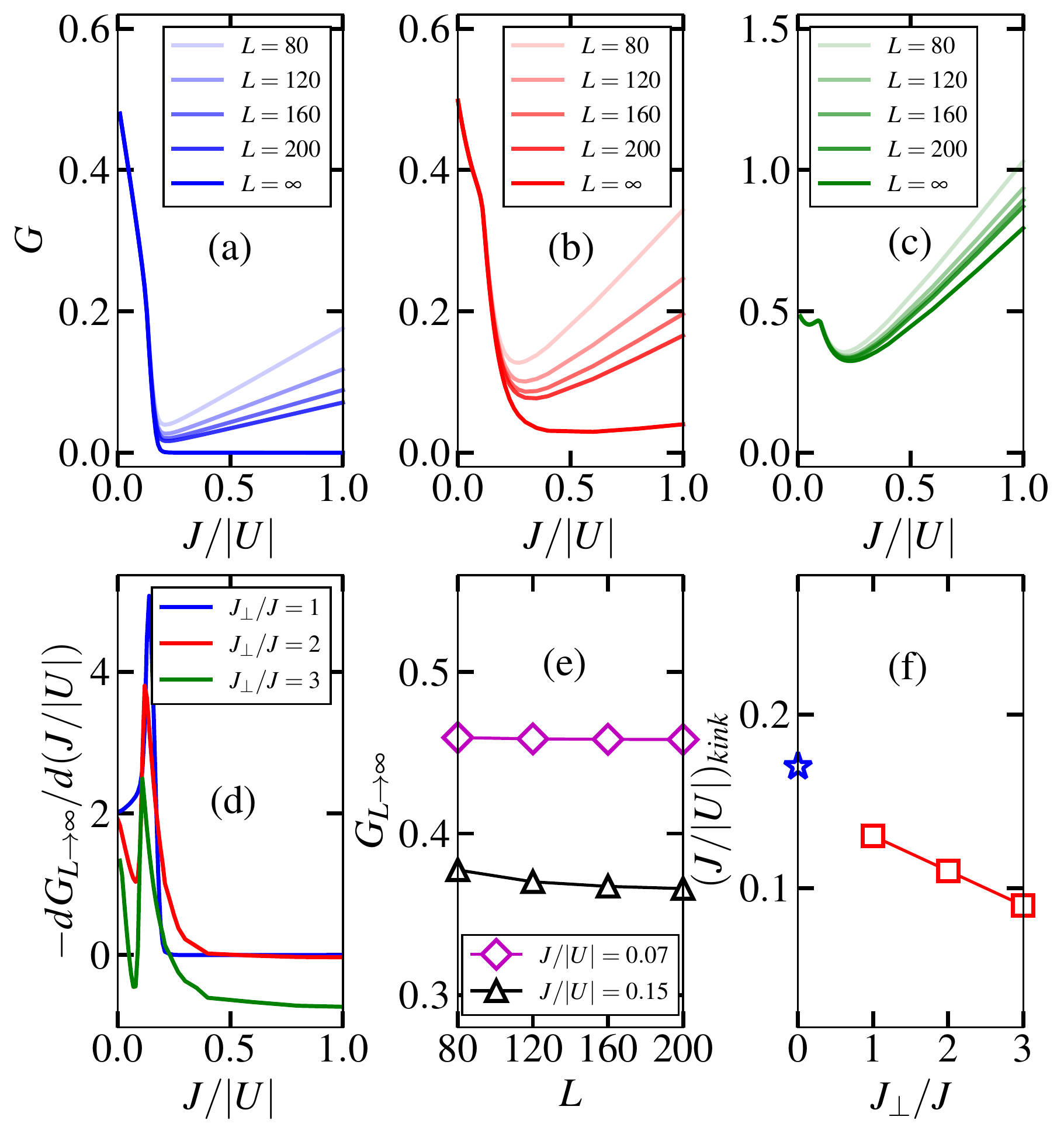}
\caption{Excitation energy gap $G$ for (a) $J_{\perp}/J=1$, (b)  $J_{\perp}/J=2$ and (c) $J_{\perp}/J=3$ at $\rho=1$ with $L=80,~120,~160,~200,~\infty$ (light to dark color). (d) The derivative of the extrapolated values of gap i.e. $G_{L\to\infty}$ with respect to $J/|U|$ for different $J_{\perp}/J$ ratios. (e) The system size dependence of $G_{L\to\infty}$ for $J/|U|=0.07$ (magenta line with diamonds) and $J/|U|=0.15$ (black line with triangles) for $J_{\perp}/J=3$. (f) Shows the kink positions in $G_{L\to\infty}$ i.e. $(J/|U|)_{kink}$ vs $J_{\perp}/J$ ratios. The blue star marks the gap-closing critical point for $J_{\perp}/J=0$ shown in Fig.~\ref{fig:hcbtbc_phases_u0tp0}.}
\label{fig:hcbtbc_phases_gap}
\end{figure}

To understand this behavior of the gap further, we plot  $G$ as a function of $J/|U|$ for $J_{\perp}/J=1,~2,~3$ in Fig.~\ref{fig:hcbtbc_phases_gap} (a-c) respectively for ladders of different lengths $L=80,~120,~160,~200$ along with the extrapolated values at the thermodynamic limit $G_{L\to\infty}$ (light to deep colour lines). It can be seen that while for $J_{\perp}/J=1$ (Fig.~\ref{fig:hcbtbc_phases_gap}(a)), $G_{L\to\infty}$ vanishes after a critical $J/|U|\sim 0.19$, for $J_{\perp}/J=2$ (Fig.~\ref{fig:hcbtbc_phases_gap}(b)) and $J_{\perp}/J=3$ (Fig.~\ref{fig:hcbtbc_phases_gap}(c)), $G_{L\to\infty}$ always remain finite as a function of $J/|U|$. However, in the latter two cases,  $G_{L\to\infty}$ first decreases, reaches a minimum and then increases, exhibiting a gapped to gapped phase transition. Note that a similar behavior is also seen in the CMFT results for a comparatively strong $J_{\perp}/J$ (see Fig.~\ref{fig:hcbtbc_bhmodel3}) although the feature of gap reaching a minimum is not clearly visible. Interestingly, in all the three cases, the gaps for all lengths exhibit a kink (change in slope) at a particular value of $J/|U|$ which is also seen in the $\mu^-$ curves shown in Fig.~\ref{fig:hcbtbc_phases_varying}. This appearance of the kink in $G$ is more prominent in the case of $J_{\perp}/J=3$. The change in slope in the gap $G$ can be seen as a discontinuity in the derivative $-d{G_{L\to\infty}}/d{(J/|U|)}$ which is shown in Fig. ~\ref{fig:hcbtbc_phases_gap}(d) for all the three case of $J_\perp/J$ considered. We find that the gap $G$ is almost system size independent up to the kink positions after which they separate from each other.  The system size independence can be seen from Fig.~\ref{fig:hcbtbc_phases_gap}(e) where we have plotted $G$ as a function of $L$ for $J/|U|=0.07$ (magenta diamonds) and $J/|U|=0.15$ (black triangles) which correspond to two arbitrary points before and after the kink location respectively for $J_{\perp}/J=3$ as an example.

\begin{figure}[t!]
\centering
\includegraphics[width=1\linewidth]{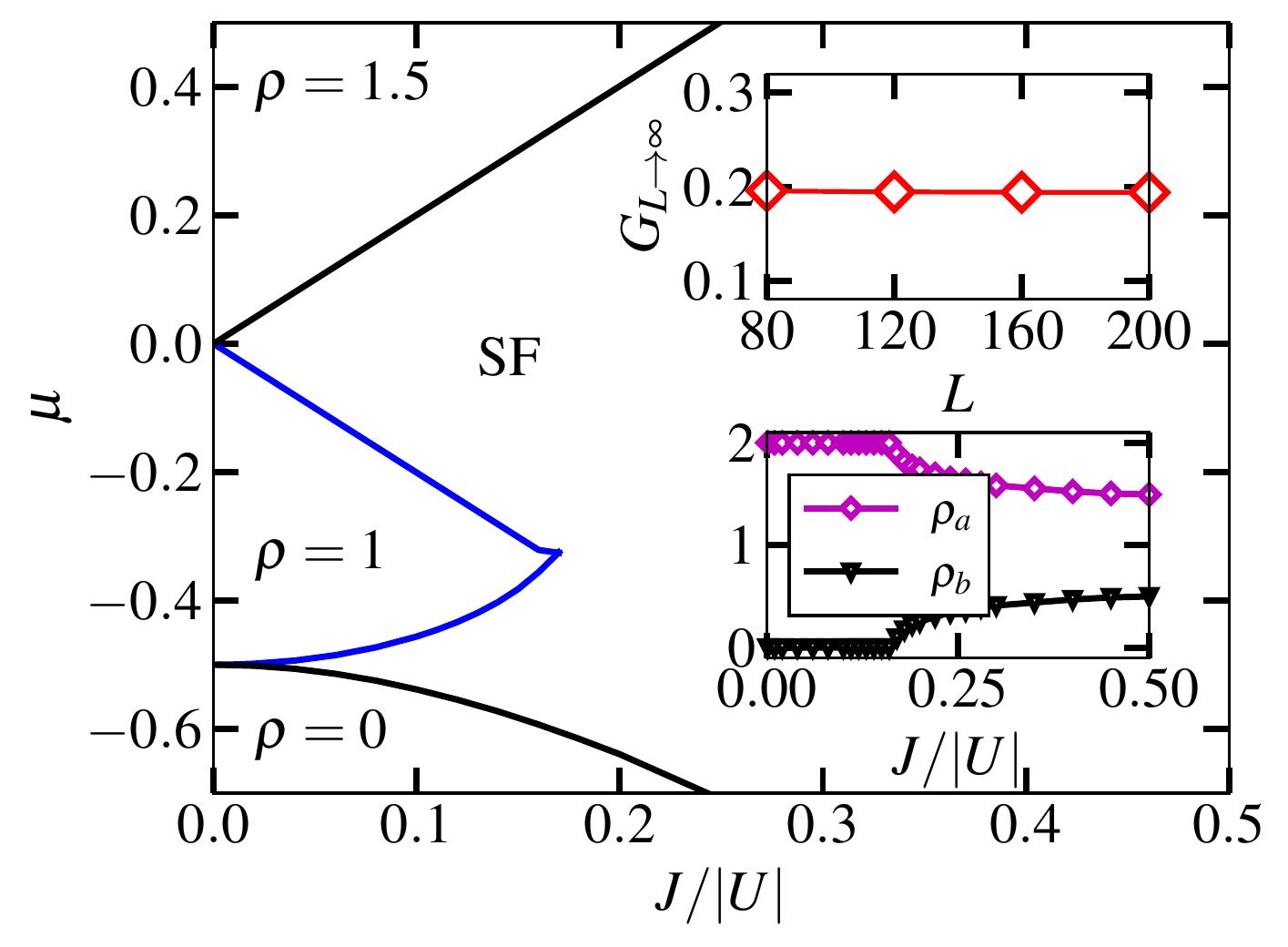}
\caption{DMRG phase diagram of the TBC-HC system for $J_{\perp}/J=0$. The phase boundaries represent the extrapolated values of $\mu^+$ and $\mu^-$ with $L=80, 120, 160$ and $200$. The upper inset shows the system size dependence of the gap $G_{L\\to\infty}$ for $J/|U|=0.12$. Lower inset shows the behavior of $\rho_a$ (magenta diamonds) and $\rho_b$ (black triangles) with respect to $J/|U|$. Here we consider the total density of the sites in the range $L/4$ to $3L/4$ on a system of size $L=200$.}
\label{fig:hcbtbc_phases_u0tp0}
\end{figure}

This particular feature of almost length independence of gap $G$ is a signature of an exact ground state which can be understood by going to $J_\perp=0$ limit. In Fig.~\ref{fig:hcbtbc_phases_u0tp0} we show the phase diagram for $J_\perp=0$ by plotting the $\mu^{\pm}$ as a function of $J/|U|$. In this situation, the system is composed of two isolated legs. In the limit of  small $J/|U|$ (i.e. strong attractive $U$), due to the TBC in leg-a, all the bosons form stable bound pairs and occupy the leg-a. At $\rho=1$, the leg-a achieves full state and leg-b remains empty which is indicated by the values $\rho_a=2$ (magenta diamonds) and $\rho_b=0$ (black down triangles) as shown in the lower inset of Fig.~\ref{fig:hcbtbc_phases_u0tp0}. Since the possible excitation in the system are in the form of bound pairs, at unit filling ($\rho=1$), the system exhibits a gap and the ground state can be called a Mott insulator of pairs~\cite{daley2009atomic} and we call it a pair-MI (PMI)
phase. 
With increase in $J/|U|$, the stable bound pairs tend to break and single particle excitation becomes energetically favorable in the system after a critical $J/|U|\sim 0.17$. Due to the single particle excitation after $J/|U| > 0.17$, the boson occupation probability in leg-b becomes finite and the system becomes gapless at $\rho=1$. At this point a sharp deviation in $\rho_a$ and $\rho_b$ from the value $2$ and $0$ respectively occurs as shown in the inset of Fig.~\ref{fig:hcbtbc_phases_u0tp0}. Note that within the gapped phase the gap remains almost independent of length (see the upper inset of Fig.~\ref{fig:hcbtbc_phases_u0tp0}) as the ground state is an almost exact state at this limit of interaction. 

When $J_{\perp}$ is turned on, the single particle excitation become feasible for smaller values of $J/|U|$ as compared to the decoupled leg limit due to the enhanced kinetic energy in the system. This behavior is indicated by the length independence of gap $G$ up to the kink position as shown in Fig.~\ref{fig:hcbtbc_phases_gap} (a-c). 
For clarity,  we have plotted the values of $(J/|U|)_{kink}$ that correspond to the kink positions for different values of $J_{\perp}/J$ in Fig.~\ref{fig:hcbtbc_phases_gap}(f) which shows a decrease in $(J/|U|)_{kink}$ with in crease in $J_\perp/J$. We have also shown the critical $J/|U|$ for $J_{\perp}=0$ (blue star) for comparison. However, contrary to the decoupled leg limit ($J_{\perp}=0$), for finite $J_{\perp}$ the gap due to the MI state does not close immediately after  $(J/|U|)_{kink}$. This is because, even though the strength of $J/|U| > (J/|U|)_{kink}$ is large enough to create single particle excitation, populating leg-b is energetically not favorable due to finite $J_{\perp}$. Further increase in $J/|U|$ (or in other words by decreasing $|U|$), the PMI phase completely melts and the probability of occupying the leg-b increases leading to a gap closing scenario. We see this behavior of vanishing of $G$ as a function of $J/|U|$ when $J_{\perp}/J=1$ as depicted  in Fig.~\ref{fig:hcbtbc_phases_varying} (blue dot-dashed lines). 
\begin{figure}[!t]
\centering
\includegraphics[width=1\linewidth]{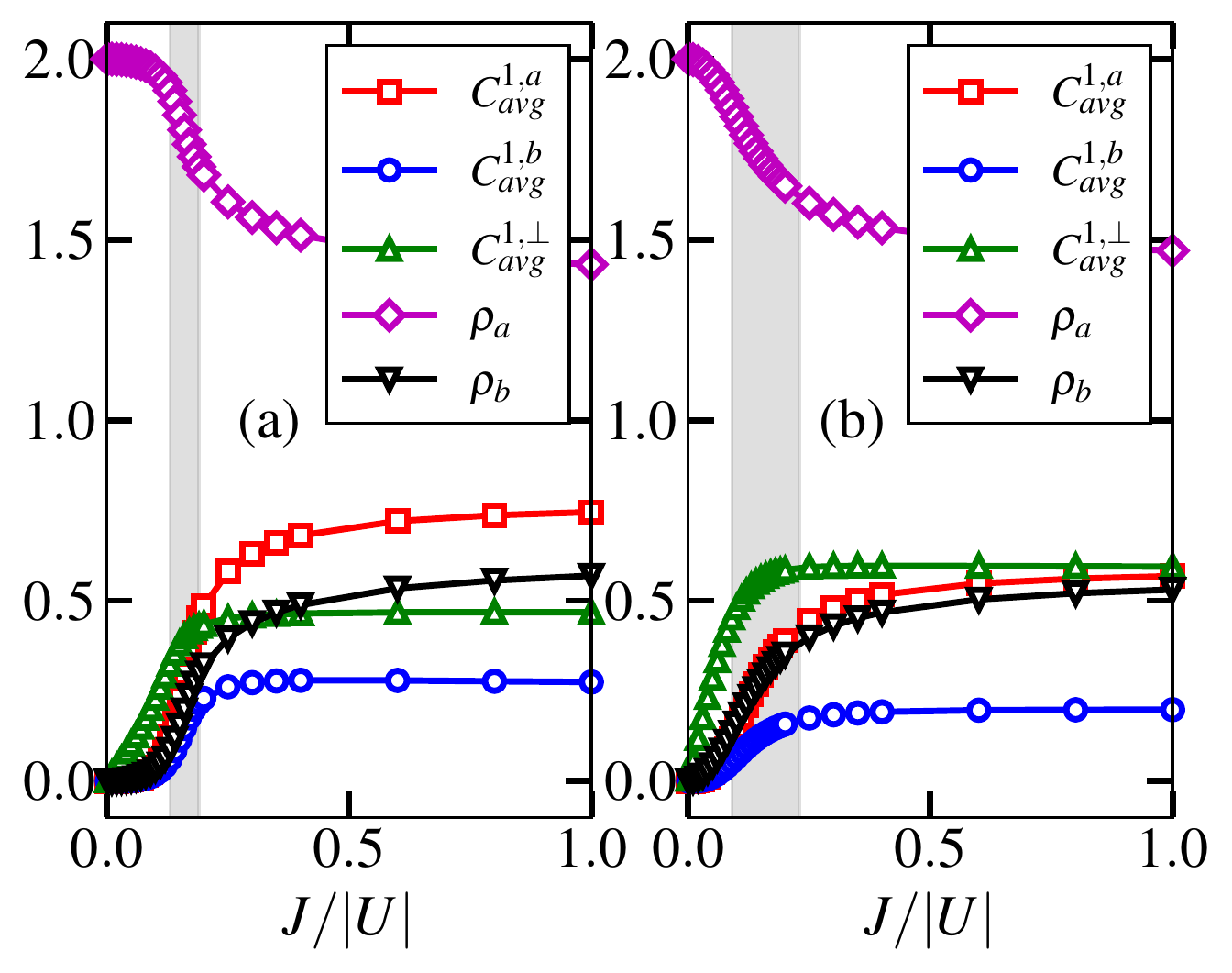}
\caption{Average nearest neighbor correlation functions and onsite densities for (a) $J_{\perp}/J= 1$ and (b) $J_{\perp}/J= 3$. Here the averaging is done by considering the lattice sites in the range $L/4$ to $3L/4$ on a system of size $L=200$. The start and end points of the shaded grey areas in both (a) and (b) represent the kink position in the gap and the point of minimum gap, respectively.}
\label{fig:dmrg_corr_avg_x}
\end{figure}

\begin{figure}[!t]
\centering
\includegraphics[width=1\linewidth]{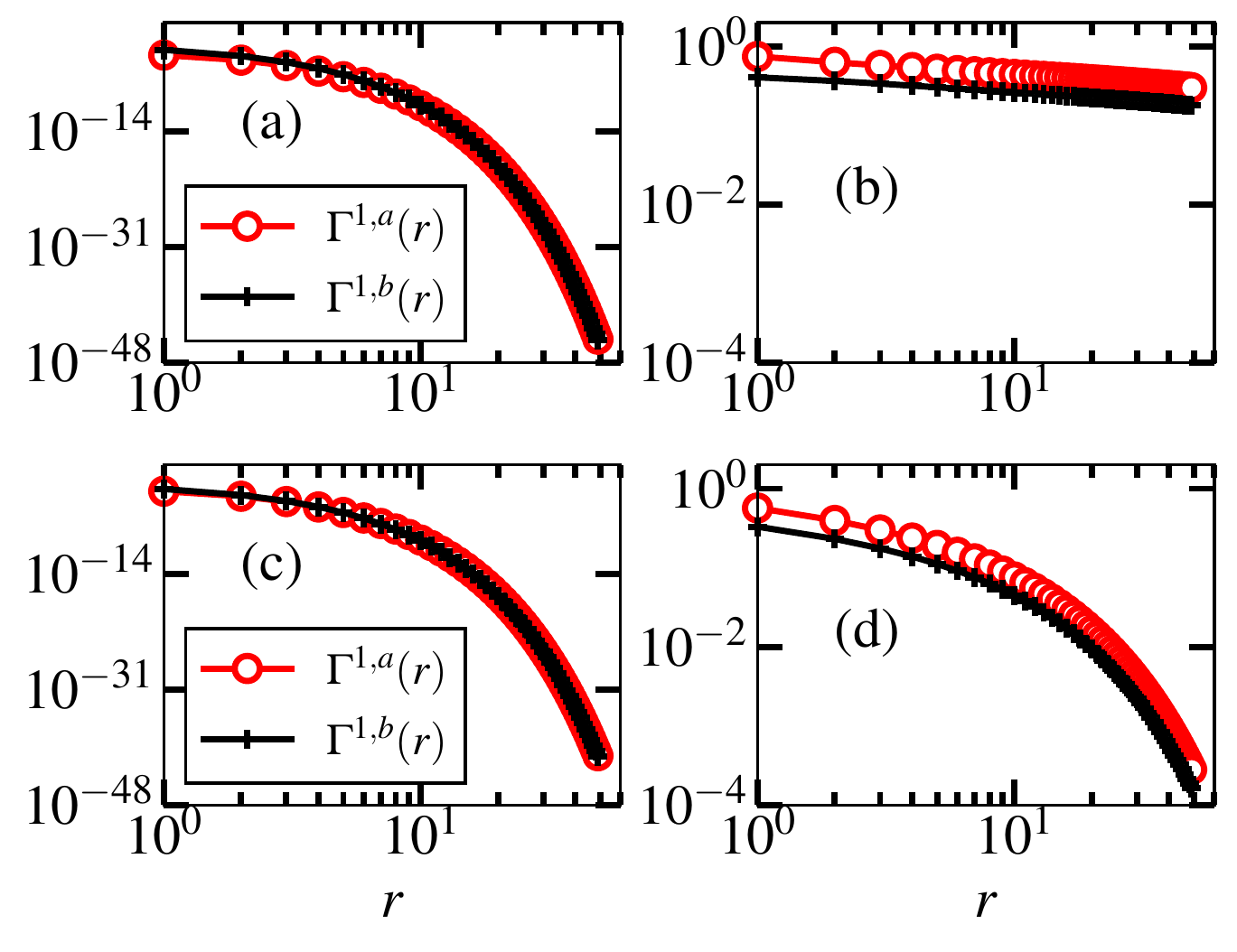}
\caption{The long-range correlations along leg-a, $\Gamma^{1,a}(r)$ and along leg-b, $\Gamma^{1,b}(r)$ plotted as a function of distance $r=|i-j|$ in log-log scale for (a) $J/|U|=0.04$ and $J_\perp/J=1$, (b) $J/|U|=1.0$ and $J_\perp/J=1$, (c) $J/|U|=0.04$ and $J_\perp/J=3$ and (d) $J/|U|=1.0$ and $J_\perp/J=3$. The correlations are plotted for the lattice sites in the range $L/4$ to $3L/4$ on a system of size $L=200$.}
\label{fig:dmrg_longcorr_various}
\end{figure}

To further understand this picture at $J_{\perp}/J= 1$, we  compute the single-particle correlation functions across all the nearest neighbor (NN) bonds along the legs  as defined in Eq.~\ref{eq:corr5} and plot their average values in Fig.~\ref{fig:dmrg_corr_avg_x} (a) as a function of $J/|U|$. It can be seen that when the system is deep in the PMI phase i.e. when $J/|U|\to 0$, all the average correlations vanish. As $J/|U|$ increases, the average NN correlations along leg-a i.e. $C^{1,a}_{avg}$ (red squares) and along leg-b i.e. $C^{1,b}_{avg}$ (blue circles) remain close to zero up to $(J/|U|)_{kink}\sim0.13$.  After the kink position in the gap, $C^{1,a}_{avg}$ and $C^{1,b}_{avg}$ start to grow due to the increase in $J/|U|$ even when the system is gapped. Note that the finite values of the average NN correlations in the gapped phase is only due to the particle-hole fluctuation and there is no long-range correlations in the system. The absence of long-range correlation along the legs can be confirmed from the behavior of the single particle correlation between a pair of sites defined as 
\begin{equation}
    \Gamma^{1,a}_{i,j}=\langle a_{ia}^\dagger a_{ja}\rangle ~{\rm and}~ \Gamma^{1,b}_{i,j}=\langle a_{ib}^\dagger a_{jb}\rangle
\end{equation}
along leg-a and leg-b respectively. In Fig.~\ref{fig:dmrg_longcorr_various}(a) we plot $\Gamma^{1,a}_{i,j}$ (red circles) and $\Gamma^{1,b}_{i,j}$ (black pluses) for $J/|U|=0.04$ and $J_\perp/J=1$. The exponential decay of the correlation function in both the legs confirm the gapped nature of the phases. With further increase in $J/|U|$, the system becomes gapless at $J/|U|\sim 0.19$ and after this critical value all the average NN correlations are finite and large. Clearly, in the gapless regime, $C^{1,a}_{avg} > C^{1,\perp}_{avg} > C^{1,b}_{avg}$. Moreover, the single particle correlation functions $\Gamma^{1,a}_{i,j}$ and $\Gamma^{1,b}_{i,j}$ exhibit power-law decay confirming the SF phase as depicted in Fig.~\ref{fig:dmrg_longcorr_various}(b) for $J/|U|=1$ and $J_\perp/J=1$. The signature of this transition can also be seen from the average particle densities $\rho_a$ (magenta diamonds) and $\rho_b$ (black down triangles) along leg-a and leg-b respectively as a function of $J/|U|$ as shown in Fig.~\ref{fig:dmrg_corr_avg_x} (a) for $J_\perp/J=1$. Initially, in the PMI state we can see $\rho_a \sim 2$ and $\rho_b \sim 0$. As $J/|U|$ increases, due to single particle excitations, $\rho_a$ ($\rho_b$) starts to decrease (increase) from the value $2~(0)$ and eventually tend to saturate at $\sim1.5~(\sim0.5)$ in the SF phase.
It is to be noted that, the average rung correlation $C^{1,\perp}_{avg}$ (green triangles) remains finite due to the finite rung hopping in the system.

On the other hand, for the ratio $J_\perp/J=2$, this gapped PMI phase extends to larger $J/|U|$ values and does not turn into an SF phase. Rather, the system enters into another gapped phase as a function of $J/|U|$. This feature is more prominent in the case of $J_\perp/J=3$ as already shown in  Fig.~\ref{fig:hcbtbc_phases_varying}. In this case, the average NN correlations behave differently as compared to the case of $J_\perp/J=1$ in the regime of large $J/|U|$,  where $C^{1,\perp}_{avg}$ dominates over the NN leg correlations $C^{1,a}_{avg}$ and $C^{1,b}_{avg}$ (Fig.~\ref{fig:dmrg_corr_avg_x}(b)). Note that even though the NN leg correlations are finite, the long-range correlations  vanish exponentially for all values of $J/|U|$. We show this behavior by plotting $\Gamma^{1,a}_{i,j}$ (red circles) and $\Gamma^{1,b}_{i,j}$ (black pluses) for two exemplary values of $J/|U|=0.04$ and $J/|U|=1.0$ in Fig.~\ref{fig:dmrg_longcorr_various}(c) and (d) respectively for $J_\perp/J=3$. In this case also, the average leg densities $\rho_a$ and $\rho_b$ saturate to values close to $1.5$ and $0.5$ respectively at large $J/|U|$ (Fig.~\ref{fig:dmrg_corr_avg_x}(b)) even though the system is gapped.

These behavior of strong rung-correlation, finite gap and  average densities approaching half-integer values are indications of an RMI phase which is exhibited by hardcore bosons  on a two-leg ladder~\cite{crepin2011}. This character of pure HCB ladder in the system of TBC-HC ladder can be understood as follows. At unit filling of the system i.e. $\rho=1$, once the leg-a having TBC
achieves density $\rho_a=1$, the remaining bosons in the leg experience hardcore constraint. Together with the leg-b, the system is now effectively an HCB ladder at half filling which exhibits an RMI phase. This confirms that the gap at large $J/|U|$ in our system is due to the RMI phase that occurs in the limit of $|U|\to 0$. As $J/|U|$ decreases or in other words $|U|$ increases, the gap tends to decrease. However, due to the extension of the PMI phase on the other extreme of $J/|U|$ the gap does not close as a function of $J/|U|$ and the system remains gapped throughout. This results in a crossover from the PMI phase to the RMI phase indicated by a minimum in the gap as can be clearly seen from Fig.~\ref{fig:hcbtbc_phases_gap}(c).

\begin{figure}[!t]
\centering
\includegraphics[width=1\linewidth]{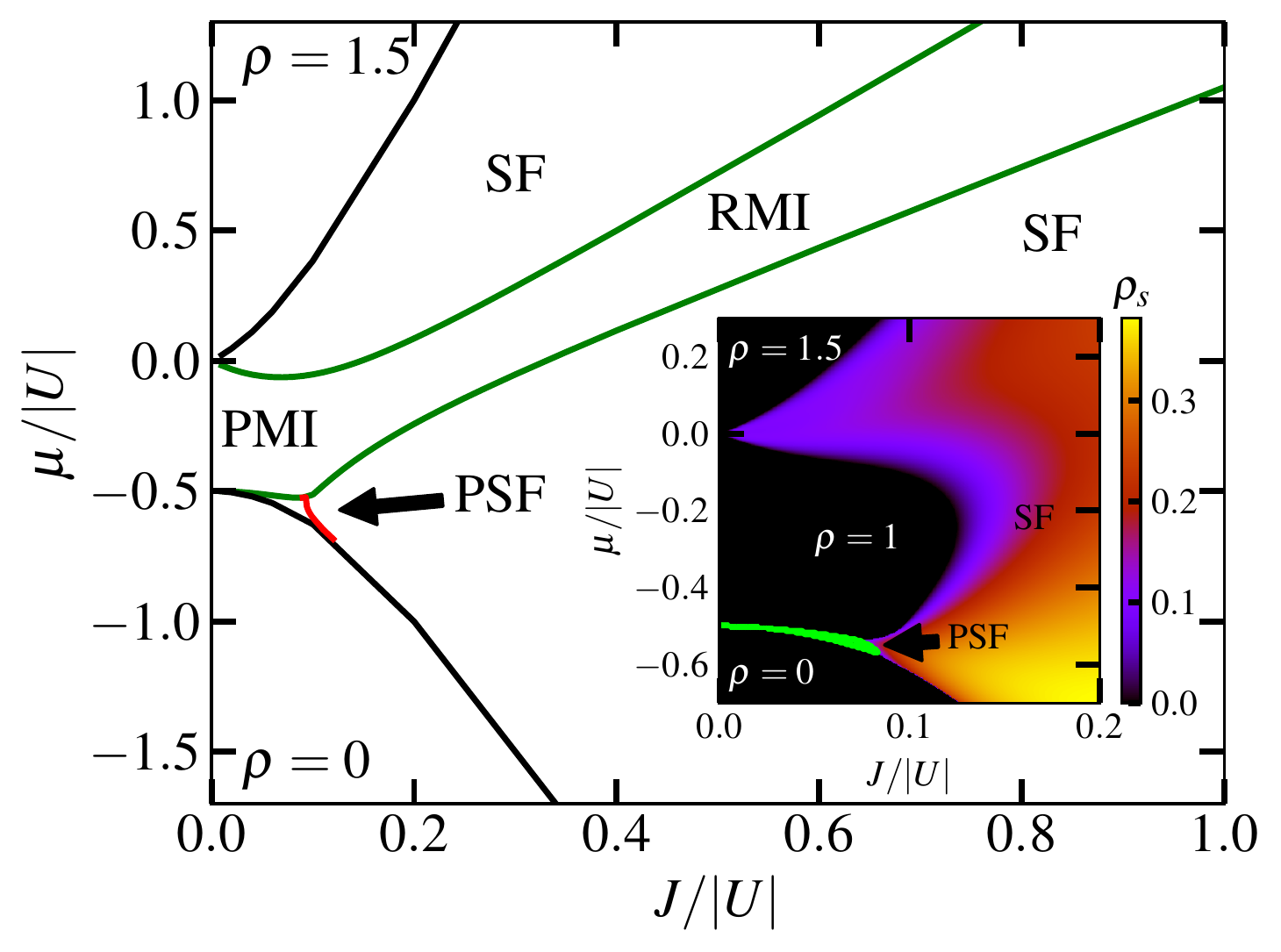}
\caption{The complete DMRG phase diagram of the TBC-HC system for $J_{\perp}/J=3$ in the $J/|U|-\mu/|U|$ plane. The green lines enclose the gapped phase(s) at $\rho=1$, the black lines represent the boundaries of the vacuum ($\rho=0$) and full ($\rho=1.5$) states and the red line denotes the boundary of the PSF phase. The phase boundaries represent the extrapolated values of $\mu^+$ and $\mu^-$  with $L=80, 120, 160$ and $200$. For comparison, the respective CMFT phase diagram is shown in the inset.}
\label{fig:hcbtbc_phases}
\end{figure}

\begin{figure}[t]
\centering
\includegraphics[width=1\linewidth]{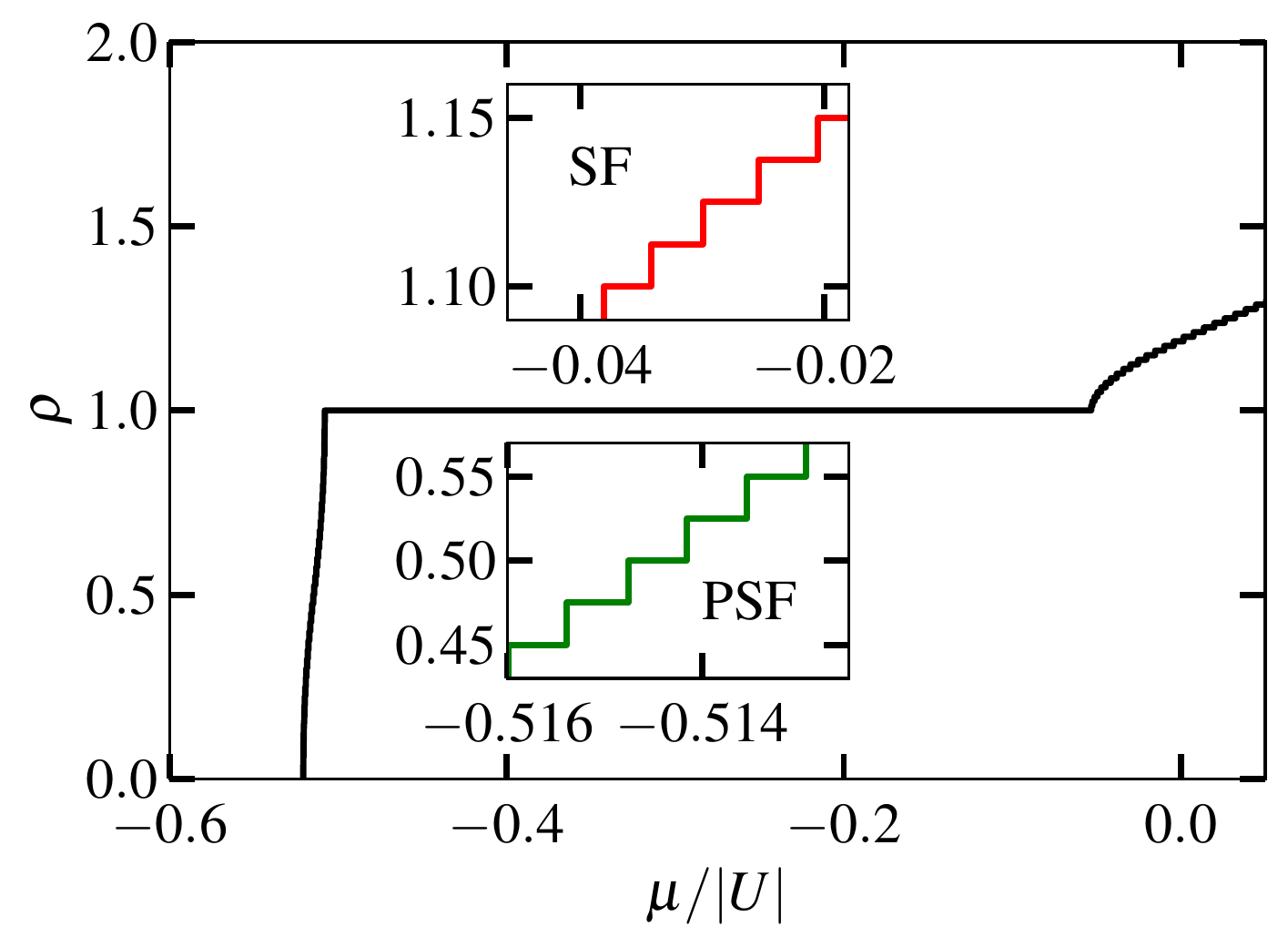}
\caption{DMRG data showing the $\rho$ vs $\mu/|U|$ plot for $J/|U|=0.04$ and $J_{\perp}/J=3$ for a system of size $L=80$. The regions above and below the plateau are enlarged in the insets for clarity which show the
signatures of the SF phase (upper inset) and the PSF phase (lower inset), respectively.}
\label{fig:hcbtbc_rho_dmrg}
\end{figure}

After having a clear idea about the gapped phases at unit filling, we focus on to understand the nature of the phases at incommensurate densities. We stress that the gapped phases in the regime of small values of $J/|U|$ for all the cases of $J_{\perp}/J$ are similar in nature. Therefore, to have a complete picture, we consider the case of $J_{\perp}/J = 3$ and depict the phase diagram of the system using the DMRG method as shown in Fig.~\ref{fig:hcbtbc_phases}.
Here the gapped region at $\rho=1$ is enclosed between the solid green lines and the empty ($\rho=0$) and full ($\rho=1.5$) states are marked by the solid black lines. Within the gapped region, the PMI and RMI phases exists at two opposite limits of $J/|U|$ and there is a crossover between them signaled by the minimum gap. The regions below and above the $\rho=1$ gapped region are found to be gapless SF phases with off-diagonal long-range order. However, in the limit of small $J/|U|$ or strong attractive interaction, we obtain a PSF phase sandwiched between the $\rho=1$ and $\rho=0$ gapped phases. The occurrence of the PSF phase is due to the favorable condition for strong bound pair formation in leg-a in the regime of small $J/|U|$. On the other hand, increasing density away from $\rho=1$ gives rise to the SF phase as the leg-b starts to get populated where hardcore constraint is imposed. This feature is confirmed by plotting the  particle density $\rho$ as a function of $\mu/|U|$ in Fig.~\ref{fig:hcbtbc_rho_dmrg} for a cut through the phase diagram of Fig.~\ref{fig:hcbtbc_phases} at $J/|U|=0.04$. The plateau at $\rho=1$ is due to the gap in the system. However, below the plateau at $\rho = 1$, the density of the system changes in steps of two particles at a time as a function of $\mu/|U|$ indicating the PSF phase (lower inset in Fig.~\ref{fig:hcbtbc_rho_dmrg}). Above the $\rho=1$ plateau, the SF phase is indicated by an increase in density in steps of one particle at a time which is depicted in the upper inset of Fig.~\ref{fig:hcbtbc_rho_dmrg}. The $\mu/|U|$ values corresponding to the beginning and end points of the two-particle jumps in density denote the two boundary points of the PSF phase. We extract the boundary of the PSF phase (red line in Fig.~\ref{fig:hcbtbc_phases}) by analysing the $\rho-\mu$ plots at different cuts through the phase diagram shown in Fig.~\ref{fig:hcbtbc_phases}.

\begin{figure}[t]
\centering
\includegraphics[width=1\linewidth]{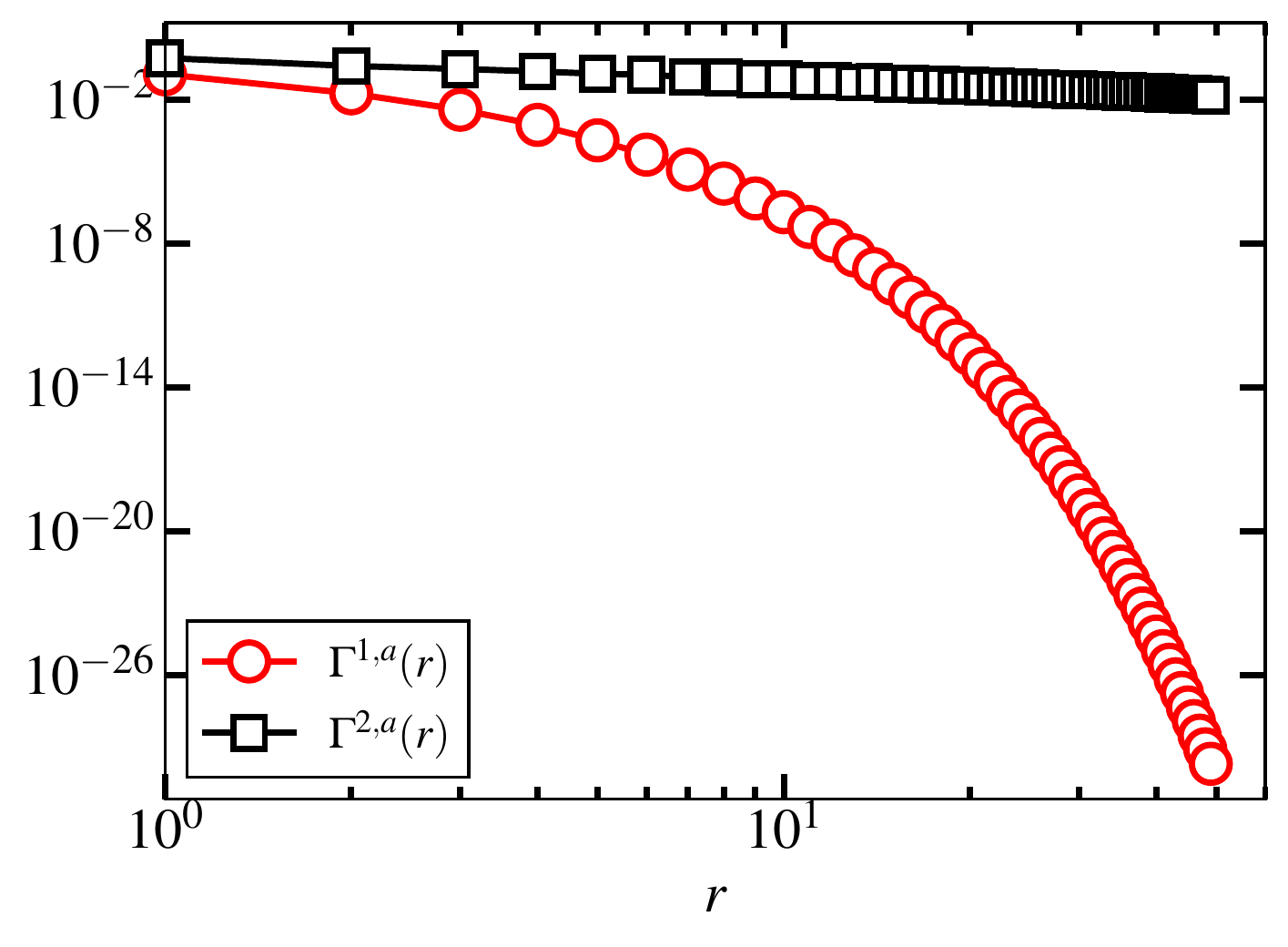}
\caption{The single-particle correlation $\Gamma^{1,a}(r)$ and the pair correlation function $\Gamma^{2,a}(r)$ for leg-a plotted as a function of distance $r=|i-j|$ in log-log scale at $\rho=0.5$ for $J/|U|=0.04$ and $J_{\perp}/J=3$. We consider the lattice sites in the range $L/4$ to $3L/4$ on a system of size $L=200$.}
\label{fig:hcbtbc_corr_dmrg}
\end{figure}

\begin{figure}[b]
\centering
\includegraphics[width=1\linewidth]{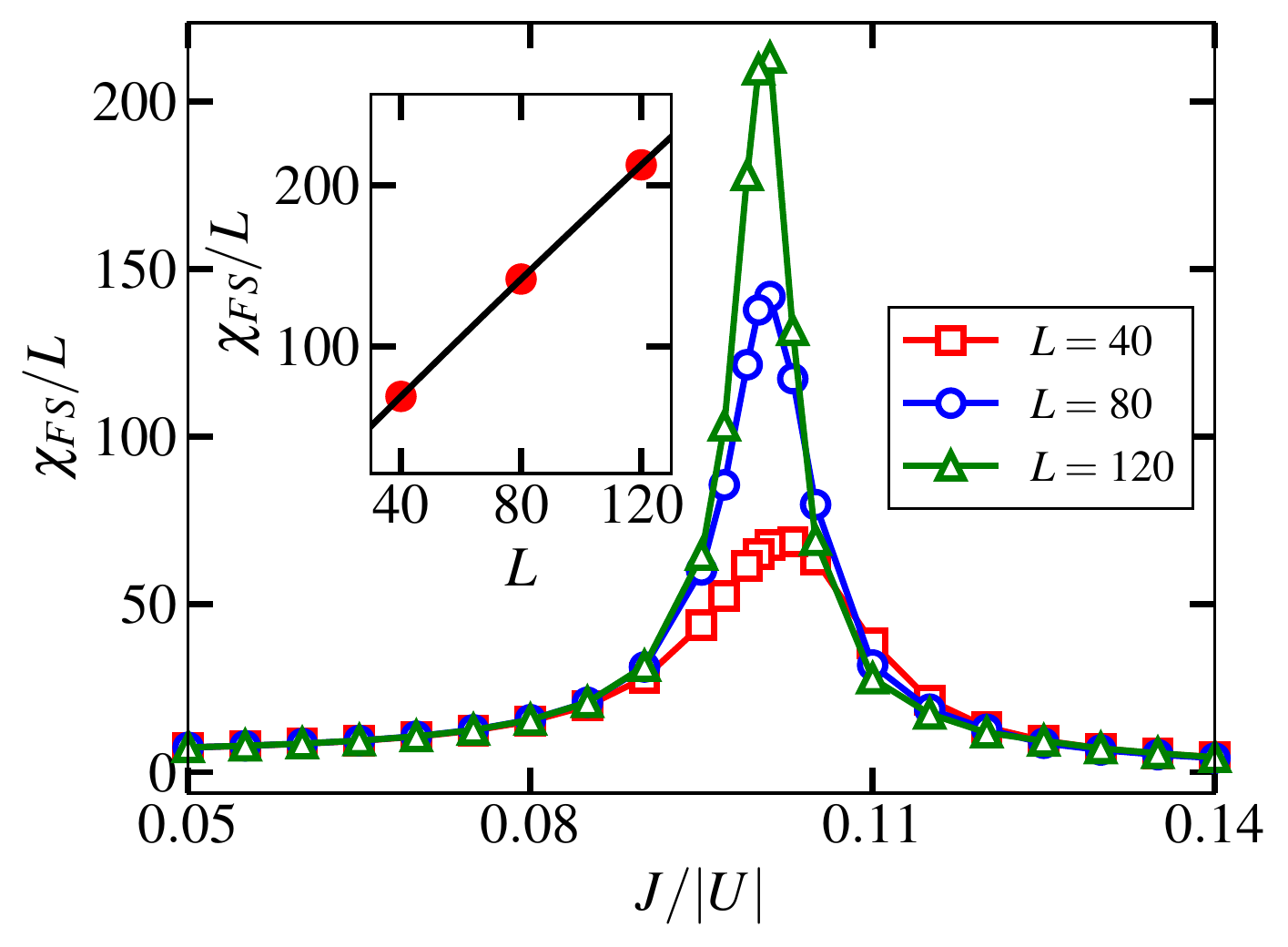}
\caption{Fidelity susceptibility $\chi_{FS}(J)$ as a function of $J/|U|$ for $L=40$ (red squares), $L=80$ (blue circles) and $L=120$ (green triangles) for $J_{\perp}/J=3$. The divergence of the peak heights for increasing system size indicates the  transition between the PSF and SF phases. Here the density of the system is fixed at $\rho=0.5$. The red points in the inset denote the peak height of the fidelity susceptibility vs system size and the black line marks the fitted line.}
\label{fig:hcbtbc_fid_dmrg}
\end{figure}

To further substantiate the PSF phase we compare the single- and pair-correlation functions as a function of distance $r=|i-j|$ along leg-a for $\rho=0.5$ and $J/|U|=0.04$ in Fig.~\ref{fig:hcbtbc_corr_dmrg}. When the system is in the PSF phase the single-particle correlation function $\Gamma^{1,a}(r)$ (red circles) decays exponentially whereas the pair-correlation function $\Gamma^{2,a}(r)$ (black squares) exhibits a power-law decay indicating the off-diagonal long-range order.
The PSF-SF phase transition points as a function of $J/|U|$ is further confirmed by looking at the behavior of the fidelity susceptibility defined as;
\begin{equation}
\chi_{FS}(J) = \lim_{(J-J^{'})\to 0}\frac{-2|\ln\langle\Psi(J)|\Psi(J^{'})\rangle|}{(J-J^{'})^2} 
\end{equation}
where $|\Psi(J^{'})\rangle$ is the ground state wavefunction for a small change $J^{'}$ in the leg hopping $J$. We plot $\chi_{FS}$ as a function of $J/|U|$ in Fig.~\ref{fig:hcbtbc_fid_dmrg} for $\rho=0.5$ and $J_\perp/J=3$. The divergence of $\chi_{FS}$ around $J/|U|\sim0.102$ and increase in peak height with increase in system size indicate the PSF-SF phase transition~\cite{gu2010fidelity,singh2018anomalous,lahiri2020correlated}. 

From the above study, it is understood that in the limit of large $J_\perp/J$  and large $J/|U|$, the TBC-HC case exhibits the RMI phase at unit filling which is a character of the HC-HC ladder at half filling. Moreover, the system also exhibits the PMI phase for small $J/|U|$ and a PSF phase below the PMI phase which are not possible to stabilize in the HC-HC ladder.
It is to be noted that that for $J_\perp/J=3$, the CMFT approach does not capture the RMI phase at higher values of $J/|U|$ which can be seen from the CMFT phase diagram shown in the inset of Fig.~\ref{fig:hcbtbc_phases}.

After obtaining the ground state properties of TBC-HC system, we now focus on to study the TBC-TBC case in the remaining part of the paper for comparison.

\subsection{The TBC-TBC system}\label{tbctbc}

\begin{figure}[!b]
\centering
\includegraphics[width=1\linewidth]{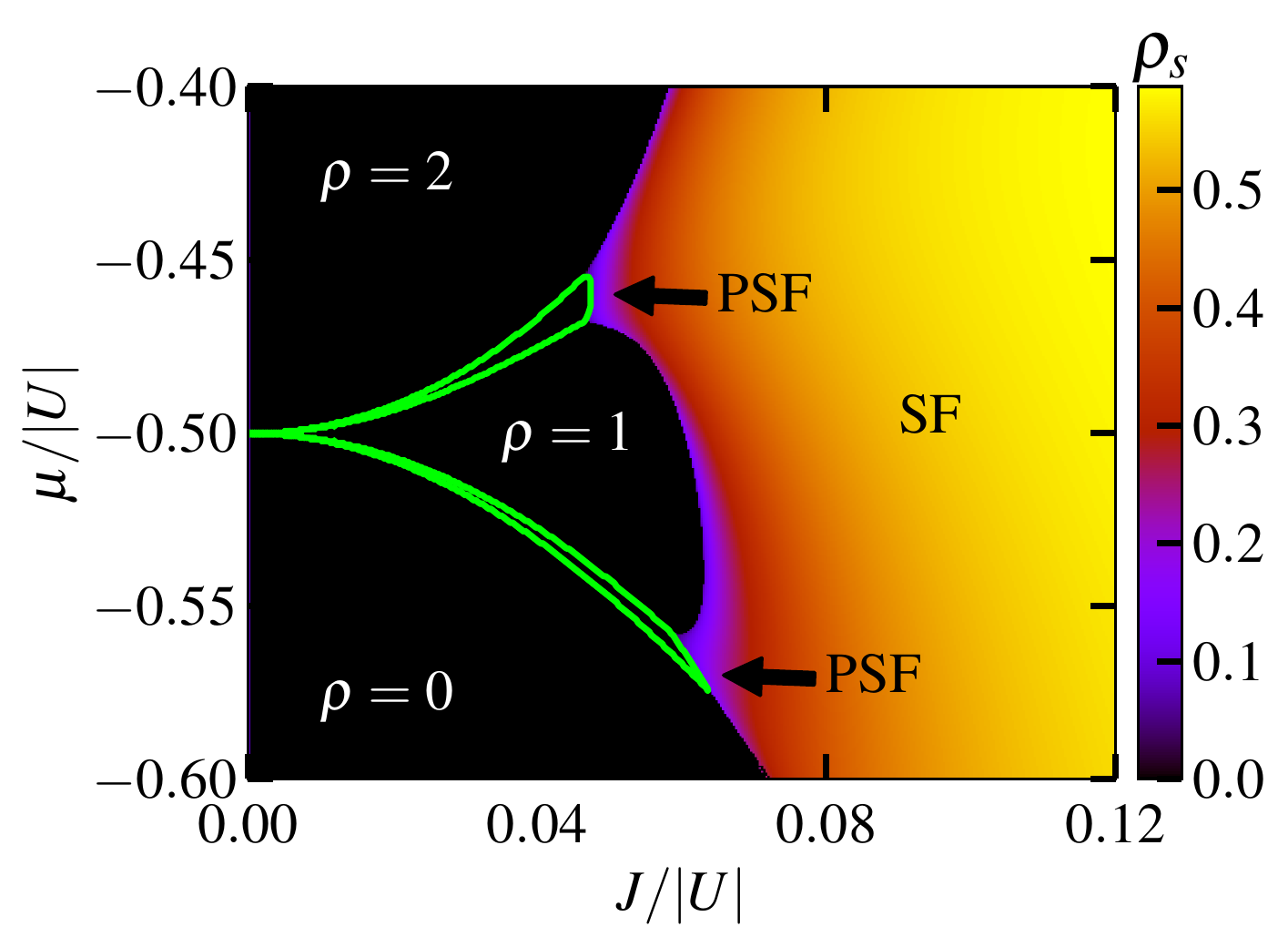}
\caption{CMFT phase diagram of the TBC-TBC ladder in the $J/|U|-\mu/|U|$ plane for $J_{\perp}/J=3$  on a $4$-site cluster. The color bar represents the values of $\rho_s$ and the solid green line denotes the boundaries of the PSF phases.}
\label{fig:tbctbc_cmft}
\end{figure}

\begin{figure}[!t]
\centering
\includegraphics[width=1\linewidth]{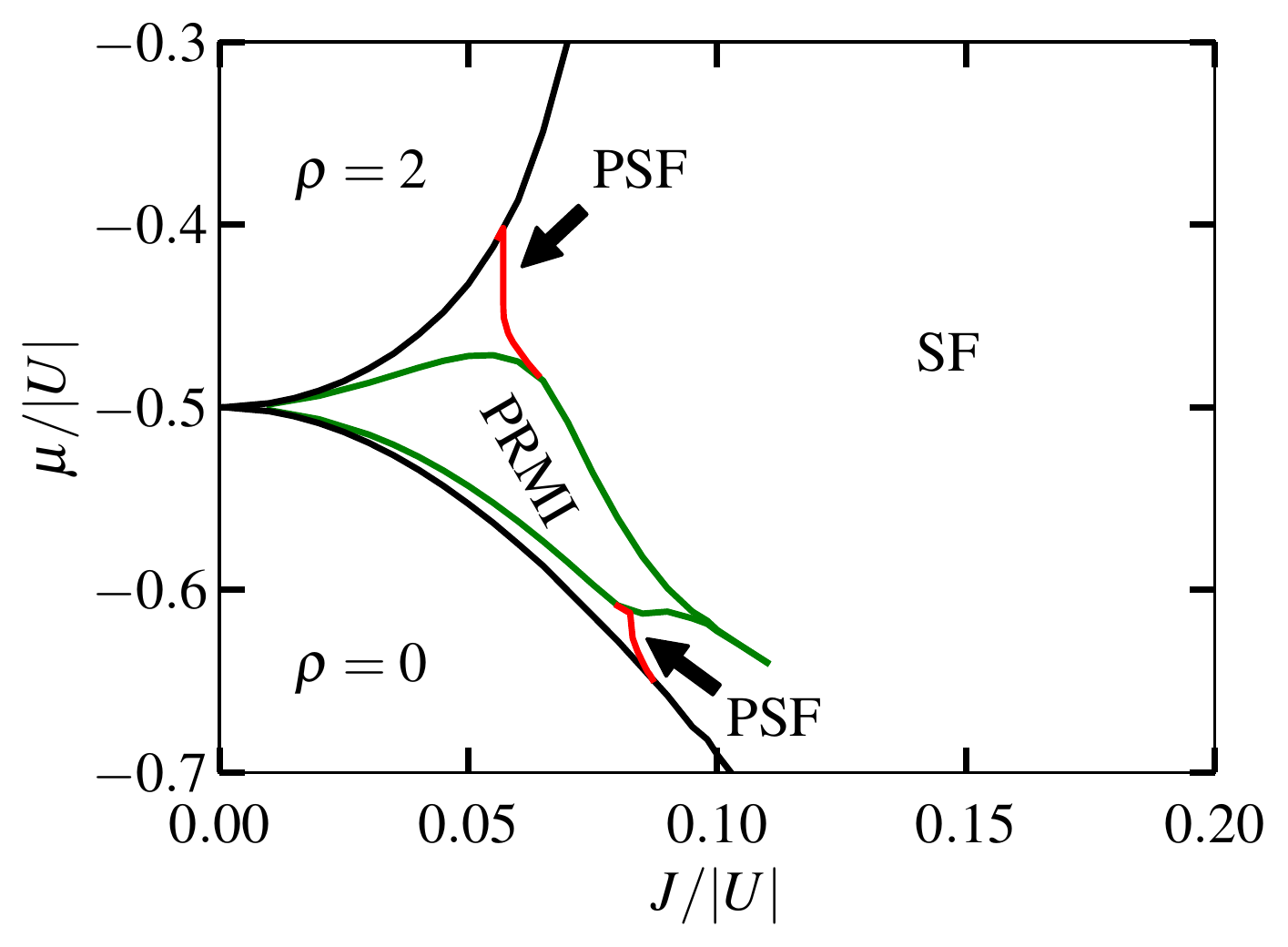}
\caption{The DMRG phase diagram for the TBC-TBC ladder in the $J/|U|-\mu/|U|$ plane for $J_{\perp}/J=3$. The green lines enclose the gapped phase at $\rho=1$, the black lines represent the boundaries of the vacuum ($\rho=0$) and full ($\rho=2$) states and the red line denotes the boundary of the PSF phases. Here the phase boundaries represent the extrapolated values of $\mu^+$ and $\mu^-$  with $L=80, 120$ and $160$.}
\label{fig:tbctbc_dmrg}
\end{figure}

We now extend our studies for a system with TBC imposed in both the legs of the ladder. In this case $|U|$ is finite in both the legs and because of its attractive nature, bound pairs tend to form in each leg for stronger values of $|U|$. In the limit of small $J/|U|$ (strong attractive $U$) the bound pairs behave like hardcore bosons due to the TBC and the system approaches the limit of HC-HC  ladder. Therefore, at $\rho=1$, due to the presence of finite $J_\perp/J$, we get the RMI  phase of the pairs which we call the pair-rung-Mott insulator (PRMI) phase where, a pair of bosons gets localized in each rung of the ladder. 

We obtain the ground state phase diagram of this system in the $J/|U|-\mu/|U|$ plane for $J_{\perp}/J=3$ by using both the CMFT and the DMRG approaches which are shown in  Fig.~\ref{fig:tbctbc_cmft} and Fig.~\ref{fig:tbctbc_dmrg} respectively. In this case  we get qualitatively similar phase diagrams using both the methods. In the phase diagrams, the gapped lobe at $\rho=1$ indicates the PRMI phase. The full and empty states are indicated by the $\rho=2$ and $\rho=0$ regions, respectively. Note that in both the phase diagrams, the gapped PRMI phase no longer survives with increase in $J/|U|$ and the system becomes a gapless superfluid. The physics behind the emergence of the PRMI-SF phase transition can be understood as follows. When $J/|U|$ is small but finite, the pairs formed due to the attractive onsite interaction can move only along the rungs since the rung hopping is dominant over the leg hopping. On increasing $J/|U|$, $J_\perp$ also increases due to the fixed ratio of the rung-to-leg hopping. Therefore, the rung localization becomes stronger for the pairs and the gap initially increases. However, further increase in $J/|U|$ starts to break the pairs into individual particles and they start moving along the legs as well. Thus, the gap starts to decrease and eventually vanishes and the system becomes a gapless superfluid as can be seen from the phase diagrams of Fig.~\ref{fig:tbctbc_cmft} and Fig.~\ref{fig:tbctbc_dmrg}. It is interesting to note that in this case the gap does not remain finite forever and the RMI phase of independent bosons is absent at $\rho=1$. Moreover, similar to the TBC-HC case here also we find the signatures of PSF phases. In this case, however, due to the TBC on both the legs, we see the appearance of the PSF phase in the regions above and below the PRMI phase in the limit of small $J/|U|$. The PSF phases are indicated by the green and and red lines in Fig.~\ref{fig:tbctbc_cmft} and Fig.~\ref{fig:tbctbc_dmrg} respectively.

\section{Conclusions}\label{conc}

In this paper, we have analyzed the ground state properties of a BH ladder with three-body constraint in one leg and hardcore constraint in the other. We have obtained the ground state phase diagrams by considering attractive onsite interaction for the bosons occupying the leg having the three-body constraint for different ratios of rung-to-leg hopping strengths. We have obtained a PMI phase to the SF phase transition at unit filling as function of the ratio between the leg-hopping and onsite interaction when the rung-to-leg hopping ratio $J_\perp/J=1$. However, when $J_\perp/J=3$, we have obtained a crossover from the gapped PMI phase to the gapped RMI phase. By moving away from unit filling, we have found the signatures of the PSF phase below the PMI region and rest of the regions are found to be in the SF phase. We have extended our studies to a system by imposing TBC on both the legs and have found a phase transition from the gapped PRMI phase to a gapless SF phase at unit filling for $J_\perp/J=3$. Moreover, we have obtained the PSF phase in either sides of the PRMI phase by moving away from unit filling. Our studies are based on the CMFT method complemented by the DMRG method. While we find that the CMFT method captures the qualitative picture of the phase diagram, it fails to provide quantitatively accurate results as compared to the DMRG method.  

Our system provides the detailed ground state properties of constrained ladder in presence of onsite interaction. Although the system under consideration is simple, it reveals interesting behavior due to the competition between different constraints, hopping strengths and onsite attractive interaction. This analysis promises to be an interesting platform to reveal some new physics in the presence of non-local interaction and also in the context of topological phase transitions in presence of interaction in the ladder systems. 

The model discussed in our analysis is purely a bosonic one where both hardcore and three-body constrained bosons are considered. Therefore, it can be of relevance to experiments involving ultracold atoms in optical ladders as well as superconducting circuits.
\clearpage

\bibliography{ladder.bib}
\end{document}